%% file: main.tex
\documentclass[aps,prb,reprint,superscriptaddress]{revtex4-2}

\usepackage{amsmath}
\usepackage{amssymb}
\usepackage{mathrsfs}
\usepackage{graphicx}
\usepackage{bbold}
\usepackage{comment}
\usepackage[dvipsnames]{xcolor}
\usepackage{bm,bbm}
\usepackage{physics}
\usepackage{tikz}
\usepackage{here}
\usepackage{url}
\usepackage{hyperref} 
\usepackage{comment}

\newcommand\beq{ \begin{eqnarray} }
\newcommand\eeq{ \end{eqnarray} }

\preprint{YITP-25-43, RIKEN-iTHEMS-Report-25}

\begin{document}

\preprint{YITP-25-43, RIKEN-iTHEMS-Report-25}

\title{ Monte Carlo study on Heisenberg model with local dipolar interaction}

\author{Etsuko Itou}
\email{itou@yukawa.kyoto-u.ac.jp}
\affiliation{Yukawa Institute for Theoretical Physics, Kyoto University, 
Kitashirakawa Oiwakecho, Sakyo-ku, Kyoto 606-8502 Japan}
\affiliation{Interdisciplinary Theoretical and Mathematical Sciences Program (iTHEMS), RIKEN,
2-1 Hirosawa, Wako, Saitama 351-0198 Japan}

\author{Akira Matsumoto}
\email{akira.matsumoto@yukawa.kyoto-u.ac.jp}
\affiliation{Yukawa Institute for Theoretical Physics, Kyoto University, 
Kitashirakawa Oiwakecho, Sakyo-ku, Kyoto 606-8502 Japan}
\affiliation{Interdisciplinary Theoretical and Mathematical Sciences Program (iTHEMS), RIKEN,
2-1 Hirosawa, Wako, Saitama 351-0198 Japan}

\author{Yu Nakayama}
\email{yu.nakayama@yukawa.kyoto-u.ac.jp}
\affiliation{Yukawa Institute for Theoretical Physics, Kyoto University, 
Kitashirakawa Oiwakecho, Sakyo-ku, Kyoto 606-8502 Japan}

\author{Toshiki Onagi}
\email{toshiki.onagi@yukawa.kyoto-u.ac.jp}
\affiliation{Yukawa Institute for Theoretical Physics, Kyoto University, 
Kitashirakawa Oiwakecho, Sakyo-ku, Kyoto 606-8502 Japan}

\begin{abstract}
Aharony and Fisher showed that non-local dipolar effects in magnetism destabilize the Heisenberg fixed point in real ferromagnets, leading to a new fixed point, called the dipolar fixed point.
The non-perturbative nature of the new fixed point, however, has not been uncovered for many decades.
Inspired by the recent understanding that the dipolar fixed point is scale-invariant but not conformal invariant, we perform the Monte Carlo simulation of the local Heisenberg-dipolar model on the lattice of $40^3$ by introducing the local cost function parameterized by a parameter $\lambda$, and study its critical exponents, which should become identical to the dipolar fixed point of Aharony and Fisher in the infinite coupling limit $\lambda = \infty$.
We find that the critical exponents become noticeably different from those of the Heisenberg fixed point for a finite coupling constant $\lambda=8$ 
(e.g. $\nu=0.601(2)(^{+0}_{-2})(^{+5}_{-4})_{\beta_c}$ in the local Heisenberg-dipolar model while $\nu=0.712(1)(^{+3}_{-0})(^{+1}_{-1})_{\beta_c}$ in the Heisenberg model), and the spin correlation function has a feature that it becomes divergence-free, implying the lack of conformal invariance.
\end{abstract}

\maketitle
\section{Introduction}
\label{sec_intro}
The origin of magnetism has been one of the greatest mysteries in nature since the days of ancient Greece and China. Even after the development of Maxwell's theory of electromagnetism and classical statistical mechanics, ferromagnetism remains a mystery. As Bohr argued in 1911 in his PhD thesis, since a static magnetic field does not work on charged particles there can be no net magnetism in statistical equilibrium. It was then Heisenberg who, in 1928, brought the brilliant idea that the origin of magnetism must be quantum mechanical~\cite{Heisenberg:1928mqa}: It is not the electromagnetic interaction but the quantum exchange effect that will cause the alignment of spins and explain ferromagnetism. His model, the Heisenberg model, has been the starting point for understanding ferromagnetism in nature from statistical mechanics. It provides the simplest statistical models to study the universal nature of Currie's phase transition with the help of the renormalization group, Monte Carlo simulation, and conformal bootstrap. This is a familiar story.

In 1973, Aharony and Fisher challenged the above conventional wisdom~\cite{Aharony_1, Aharony_2}. They pointed out that the long-range electromagnetic dipolar exchange between magnetization vectors, rather than the short-range quantum mechanical exchange effect of Heisernerbg, should lead to a relevant effect in the renormalization group sense although it is typically too tiny to explain the ferromagnetism itself. Accordingly, they further argue that it can drastically change the nature of Currie's phase transition because in the theory of renormalization group, it is not the smallness of the bare parameter but whether it is relevant or irrelevant will determine the fixed point and nature of the phase transition. The salient features of the new renormalization group fixed point caused by the long-range electromagnetic dipolar exchange effect, which we call dipolar fixed point, are (1) it demands the magnetization vector fields to be constrained to be transverse, (2) as a consequence it is scale-invariant but not conformal invariant, (3) the critical exponents must be different from those of the Heisenberg fixed point.

There are some experimental verifications of (1) and (2). Certain ferromagnetic materials such as Europium compounds EuO and EuS experimentally show the spin correlation functions with suppressed transverse components~\cite{J.Kotzler_1986}, which is not only a clear signal of (1) but also that there is no conformal invariance as argued in Ref.~\cite{Gimenez-Grau:2023lpz}. Regarding (3), the measured critical exponents of EuO and EuS are close to those of the Heisenberg model~\cite{PhysRevB.14.4908}, and the difference, if any, is yet to be seen. 

We here point out that there have not been as many theoretical works to determine the critical exponents of the dipolar fixed points as those of the Heisenberg fixed points. In addition to the one or two-loop results in the original paper by Aharony and Fisher, we have only three-loop computations in Ref.~\cite{2022NuPhB.98515990K} and the functional renormalization group analysis in Ref.~\cite{Nakayama:2023wrx}. The situation should be in stark contrast with the case of the Heisenberg model, where the competition among the renormalization group~\cite{Guida:1998bx,2001JMP....42...52J}, Monte Carlo simulations~\cite{Hasenbusch:2011zwv, Hasenbusch:2020pwj}, and conformal bootstrap~\cite{Kos:2015mba, Kos:2016ysd, Chester:2020iyt} are very severe, now culminating up to more than six digits. We should emphasize here that the dipolar fixed point is not conformal invariant and the conformal bootstrap is not applicable.

In this paper, we present the first Monte Carlo predictions of critical exponents of a local Heisenberg-dipolar model by performing the simulation on a cubic lattice of size up to $40^3$. In our work, we impose the dipolar constraint not as a consequence of the long-range interaction, but as a consequence of the local energy cost. The strength of the local energy cost is measured by a parameter $\lambda$ in our notation. The $\lambda=0$ corresponds to the Heisenberg fixed point and $\lambda = \infty$ corresponds to the dipolar fixed point of Aharony and Fisher. The merit of the local Heisenberg-dipolar Hamiltonian is that we can evaluate the updating algorithm of the Monte Carlo simulation quickly. The demerit may be that taking larger $\lambda$ demonstrates longer auto-correlation time and slower thermalization. Indeed, these will hinder us from taking the $\lambda = \infty$ limit directly, but we study the theory up to $\lambda = 8$ with numerical extrapolation to $\lambda = \infty$ in mind.

\begin{figure}[h]
    \centering
    \includegraphics[width=0.8\columnwidth]{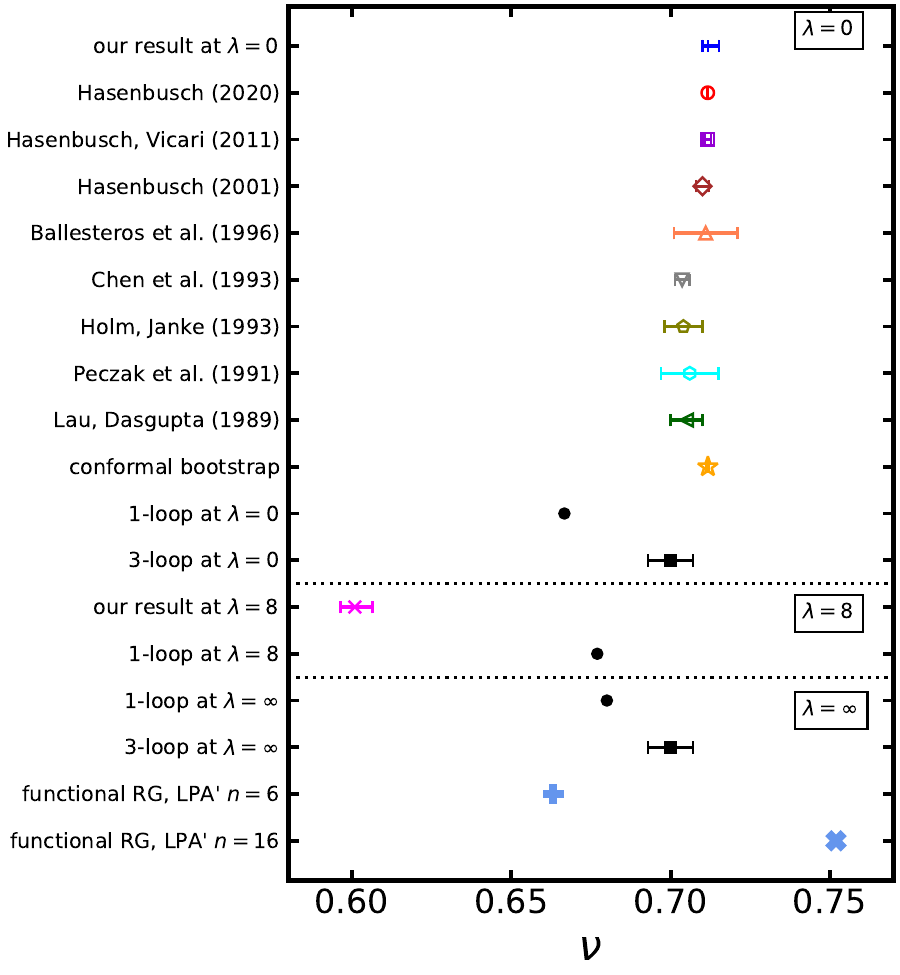}
    \caption{\label{fig_nu_comp}
    Comparison plot on the critical exponent $\nu$ (for $\lambda=0$ and $8$) obtained in this paper and the previous studies 
    by the Monte Carlo~\cite{
    Hasenbusch:2020pwj, Hasenbusch:2011zwv, Hasenbusch:2000ph, 
    Ballesteros:1996bd, Chen:1993zz, Holm:1993zz, Peczak:1991zz, Lau:1989zz} (colored open symbols), 
    the conformal bootstrap~\cite{Chester:2020iyt} (star-yellow symbol), 
    the $\varepsilon$-expansion (filled symbols), 
    and the functional renormalization group~\cite{Nakayama:2023wrx} (cross-blue-gray symbols).
    The filled circle-black symbols depict the 1-loop results of the $\varepsilon$-expansion 
    obtained for $\lambda=0$ and $\infty$ in Ref.~\cite{Aharony_1} and for $\lambda =8$ in this paper 
    whereas the filled square-circle ones are the 3-loop results taken from Ref.~\cite{2022NuPhB.98515990K}.
    The two results of the functional renormalization group are obtained with the LPA' approximation, 
    truncating the potential at the order $n=6$ or $16$~\cite{Nakayama:2023wrx}.}
\end{figure}

Our main finding is that the measured critical exponents of the dipolar fixed points can be very different from those of the Heisenberg fixed points: 
$\nu=0.601(2)(^{+0}_{-2})(^{+5}_{-4})_{\beta_c}$ and $\eta=0.132(8)(^{+0}_{-4})(^{+12}_{-11})_{\beta_c}$ at $\lambda=8$, 
which should be compared with $\nu=0.712(1)(^{+3}_{-0})(^{+1}_{-1})_{\beta_c}$ and $\eta=0.0318(27)(^{+10}_{-7})(^{+27}_{-28})_{\beta_c}$ at $\lambda=0$.
In Figure~\ref{fig_nu_comp}, we compare our results of $\nu$ with the previous Monte Carlo studies~\cite{
Hasenbusch:2020pwj, Hasenbusch:2011zwv, Hasenbusch:2000ph, 
Ballesteros:1996bd, Chen:1993zz, Holm:1993zz, Peczak:1991zz, Lau:1989zz} 
as well as the conformal bootstrap~\cite{Chester:2020iyt}, the $\varepsilon$-expansion~\cite{2022NuPhB.98515990K}, 
and the functional renormalization group~\cite{Nakayama:2023wrx}.
As for $\lambda=0$, our result is almost consistent with the other numerical works, while for $\lambda=8$ it shows a clear difference from the one at $\lambda=0$. Thus, it strongly suggests that the dipolar fixed point realized at $\lambda = \infty$ would be in a different universality class from the Heisenberg fixed point.

We also show that the correlation functions have suppressed transverse components, reflecting the dipolar constraint. This is a clear signal that the dipolar constraints defined by a local description work well in our Monte Carlo simulations and a lack of conformal invariance in correlation functions emerges.

The rest of the paper is organized as follows.
In Section~\ref{sec_theory}, we present a field theoretical description of the Heisenberg model with the dipolar constraint
and then discuss the predicted correlation function and the loop calculation of critical exponents.
Furthermore, we also present a local description of the model, which gives a rather suitable formulation for lattice calculations.
Then, we define the lattice Hamiltonian and observables for the simulation in Section~\ref{sec_lattice_formula} and explain the details of the simulation in Section~\ref{sec_simulation_detail}. 
The effectiveness of the dipolar-constraint term is also discussed.
In Section~\ref{sec_result}, we first explain the basic idea of our finite-scaling analysis and then show the main results of the critical temperature and exponents.
The characteristic behavior of correlation functions near the critical temperature is discussed in Section~\ref{sec_cf_result}. 
Section~\ref{sec_summary} is devoted to the summary and discussion.
In Appendix~\ref{sec:beta-fn}, we present the result of the renormalization group analysis.
In Appendix~\ref{sect_HS_trsf}, we discuss the Hubbard–Stratonovich transformation of the model.
In Appendix~\ref{sec_chi_peak}, the fitting results of the magnetic susceptibility for determining its peak temperature are shown.
The fitting results for interpolating $\beta$-dependent observables are shown in Appendix~\ref{sec_fit_result_interpolate}.
We discuss the system size and temperature dependences of the correlation function in Appendix~\ref{sec_l_dep_cf} and~\ref{sec_off_peak_cf}, respectively.

\section{Heisenberg-dipolar model}
\label{sec_theory}

\subsection{Hamiltonian in the continuum theory}
The Hamiltonian of the Landau-Ginzburg-Wilson theory describing three-dimensional ($d=3$) isotropic ferromagnet (i.e. the same universality class as the Heisenberg model) is
\begin{equation}
    H_{\mathrm{Heisenberg}} = \int\ d^3x \left(\frac{1}{2}\partial_i\phi_j\partial_i\phi_j + \frac{t}{2}(\phi_i\phi_i) + \frac{u}{4}(\phi_i\phi_i)^2\right) ,
\end{equation}
where $\phi_i (i=1,2,3)$ is three-component scalar field.
\footnote{In this paper, we take the sum over spatial index i.e. $i$ and $j$ that appear repeatedly.} 
By fine-tuning $t$, we reach the infrared (IR) fixed point that describes the Heisenberg model at criticality.

Motivated by the long-range electro-magnetic dipolar exchange effect of the magnetization vector, Aharony and Fisher introduced the dipolar-interaction term:
\begin{equation}
    V_{\mathrm{dip}} = v\int\  d^3xd^3 y U_{ij}(x-y)\phi_i(x)\phi_j(y),
\end{equation}
with
\begin{equation}
    U_{i j}(x)=-\partial_{x_{i}} \partial_{x_{j}} \frac{1}{|x|}=\frac{\delta_{i j}-3 \hat{x}_{i} \hat{x}_{j}}{|x|^{3}}.
\end{equation}
In momentum space, the dipolar interaction can be written as
\begin{equation}
    V_{\mathrm{dip}}=4 \pi v \int \frac{d^{3} q}{(2 \pi)^{3}} \frac{q_{i} q_{j}}{q^{2}} \phi_{i}(q) \phi_{j}(-q).
    \label{eq:Vdip}
\end{equation}
The symmetry of the original Hamiltonian $H_{\mathrm{Heisenberg}}$ was $O(3)\times O(3)$, but it is broken down to $O(3)$ by this interaction.
In the $v\to\infty$ limit, the dipolar term forces the constraint on $\phi_i$:
\begin{equation}
    \partial_i \phi_i(x) = 0.
    \label{eq:dip_constraint}
\end{equation}
From the dimensional analysis, this interaction is relevant. In the IR limit, we should arrive at a new fixed point called the dipolar fixed point. The fixed point has the peculiar property that it is scale-invariant but not conformal invariant~\cite{Gimenez-Grau:2023lpz}. 

Were it conformal invariant, the two-point functions of $\phi_i$ should become
\begin{equation}
    \ev{\phi_i(x)\phi_j(0)} = \frac{A}{|x|^{2\Delta_{\phi}}}\qty(\delta_{ij} - 2\frac{x_i x_j}{x^2}),
    \label{eq:corr_conf}
\end{equation}
where $\Delta_\phi$ is the scale dimension of $\phi_i$.
On the other hand, the constraint \eqref{eq:dip_constraint} demands
\begin{equation}
    \ev{\phi_i(x)\phi_j(0)} = \frac{A}{|x|^{2\Delta_{\phi}}} \qty(\delta_{ij} - \alpha\frac{x_i x_j}{x^2}),
    \label{eq:corr-fn}
\end{equation}
with
\begin{equation}
    \alpha = \frac{2\Delta_{\phi}}{2\Delta_{\phi} - (d-1)}.
    \label{eq:def-alpha}
\end{equation}
The perturbative computation suggests that the value of the anomalous dimension is small; $\gamma_\phi \approx 0.0115 \varepsilon^2$~\cite{Aharony_2} with $\Delta_\phi = \frac{d-2}{2} + \gamma_\phi$ in $4-\varepsilon$ dimensions, and the perturbative fixed point cannot be conformal invariant. We attempt the non-perturbative computation of $\gamma_\phi$ from the lattice Monte Carlo simulation in this work.

For later discussions of numerical results of the correlation functions, let us note that Eq.~\eqref{eq:corr-fn} indicates that the functional form of the two-point function depends on the combination of the component $\phi_i$ and the spatial direction of correlation $x_i$ in a specific manner.
For instance, if we take $\vec{x}=(x,0,0)$, then the second term of $\qty(\delta_{ij} - \alpha\frac{x_i x_j}{x^2})$ gives $(1-\alpha)$ for $i=j=x$ component, while it does $1$ for $i=j=y$ (or $z$) component. 

Although it is not directly relevant to our work, we note that there is a hidden shift symmetry in this model.
To see this let us use the Lagrange multiplier formalism to impose the constraint \eqref{eq:dip_constraint} within the local Hamiltonian. With the Lagrange multiplier field $U$, we add the constraint term $H_{\text{contstraint}} = \int d^3x\, U \partial_i \phi_i$. Physically, we may interpret $U$ as a magnetic potential.
This added Hamiltonian has a shift symmetry, $U \to U + \text{const.}$, and we can use it to prove certain non-renormalization properties of the scale-invariant but non-conformal field theory~\cite{Gimenez-Grau:2023lpz}. 
Introducing the Lagrange multiplier on the Monte Carlo simulation, however, is difficult, so in the next subsection we further add the $U^2$ term to make the constraint into a (positive) cost term in the Hamiltonian at the expense of losing the shift symmetry.

\subsection{The local description}
It is technically challenging to simulate models with long-range interactions like in Eq.~\eqref{eq:Vdip}, so we reformulate it to a local description. 
For this purpose, instead of the long-range interaction term, we add a local term:
\begin{equation}
     H_{\mathrm{local}} = \int\ d^3x \left(\frac{1}{2}\partial_i\phi_j\partial_i\phi_j + \frac{t}{2}(\phi_i\phi_i) + \frac{u}{4}(\phi_i\phi_i)^2
     + \lambda(\partial_i \phi_i)^2 \right),
     \label{eq:local_discription}
\end{equation}
This will be called the local Heisenberg-dipolar Hamiltonian.

By fine-tuning $t$, we will reach a non-trivial fixed point. The renormalization group analysis at one-loop is carried out in Appendix~\ref{sec:beta-fn}. We claim that the fixed point with $\lambda = \infty$ is the same fixed point as the one for the non-local dipolar Heisenberg model by Aharony and Fisher.

At the one-loop order, we find $\lambda$ can take any value and the RG eigenvalues are given by 
\begin{equation}
    \frac{1}{\nu} = y_t = 2 -\frac{9 \lambda  (\lambda +1)+3}{\lambda  (17 \lambda +18)+6}\varepsilon.
    \label{eq:1-loop}
\end{equation}
In the \( \lambda \to 0 \) limit, the result, $y_t=2- \varepsilon/2$, reproduces the one of the \( O(4) \)-vector model.
On the other hand, in the \( \lambda \to \infty \) limit, it agrees with the Aharony-Fisher result, $y_t = 2- 9\varepsilon/17$~\cite{Aharony_1}.  Beyond two loops, $\lambda$ should be either $0$ or $\infty$ at the fixed point but this one-loop estimate with variable $\lambda$ will be useful to discuss the comparison with the Monte Carlo simulations where $\lambda$ takes a finite value.

For reference, we note the relation between the scaling dimensions of operators and the critical exponents:
\begin{equation}
    \Delta_\phi = \frac{-2 + d}{2} + \frac{\eta}{2},
    \label{eq_delta_and_eta}
\end{equation}
\begin{equation}
    \Delta_{\phi^2} = d - \frac{1}{\nu},
\end{equation}
\begin{equation}
    \gamma = \nu(2-\eta).
    \label{eq_gamma_eta_rel}
\end{equation}

\section{Lattice formula}
\label{sec_lattice_formula}
Let us introduce our lattice Hamiltonian  defined on the three-dimensional lattice:
\begin{equation}
    H=\sum_{n\in\Lambda}\left[J\sum_{i=1}^{3}\vec{S}_{n}\cdot\vec{S}_{n+\hat{i}}
    +\lambda\left(\vec{\nabla}\cdot\vec{S}_{n}\right)^{2}\right],
    \label{eq_lattice_model_calc}
\end{equation}
where $\Lambda$ represents a set of lattice sites of the size $|\Lambda|=\sum_{n\in\Lambda}1$.
The spin variable on the site $n$ is described by the 3-component vector $\vec{S}_n = (S_n^x,S_n^y,S_n^z)$, 
where the normalization condition $\vec{S}_n \cdot \vec{S}_n = 1$ is imposed.
The difference operator on the lattice is defined by 
\begin{equation}
    \vec{\nabla}f_{n} = 
    \left( f_{n+\hat{1}}-f_{n},\ f_{n+\hat{2}}-f_{n},\ f_{n+\hat{3}}-f_{n} \right).
    \label{eq_diff_lattice}
\end{equation}
This Hamiltonian is the Heisenberg model, adding the local cost function for the dipolar constraint~\footnote{Our normalization of $\lambda$ when $J=-1$ here is chosen such that it directly corresponds to the continuum field-theory description in Section~\ref{sec_theory} within the mean-field theory analysis based on the Hubbard-Stratonovich transformation. See Appendix~\ref{sect_HS_trsf}.}.

We can rewrite the Hamiltonian as follows;
\begin{align}
    H = & 2\lambda\left|\Lambda\right|+\sum_{n\in\Lambda}\sum_{i=1}^{3}\vec{S}_{n}\cdot M_{i} \vec{S}_{n+\hat{i}}\nonumber \\
    & +\sum_{n\in\Lambda}\sum_{ij=12,23,31}\vec{S}_{n}\cdot M_{ij}\vec{S}_{n-\hat{i}+\hat{j}}
    +\sum_{n\in\Lambda}\vec{S}_{n}\cdot M_{\mathrm{self}}\vec{S}_{n}.
    \label{eq_H_by_M}
\end{align}
Here we introduced the spin interaction matrices $M_{i}$ for the nearest neighbor sites, 
\begin{align}
    &M_{1},M_{2},M_{3} \notag\\
    &=
    \begin{pmatrix}
        J-2\lambda & 0 & 0\\
        -2\lambda & J & 0\\
        -2\lambda & 0 & J
    \end{pmatrix},
    \begin{pmatrix}
        J & -2\lambda & 0\\
        0 & J-2\lambda & 0\\
        0 & -2\lambda & J
    \end{pmatrix},
    \begin{pmatrix}
        J & 0 & -2\lambda\\
        0 & J & -2\lambda\\
        0 & 0 & J-2\lambda
    \end{pmatrix};
\end{align}
the interaction $M_{ij}$ for the next nearest neighbor sites, 
\begin{equation}
    M_{12},M_{23},M_{31}=
    \begin{pmatrix}
        0 & 2\lambda & 0\\
        0 & 0 & 0\\
        0 & 0 & 0
    \end{pmatrix},
    \begin{pmatrix}
        0 & 0 & 0\\
        0 & 0 & 2\lambda\\
        0 & 0 & 0
    \end{pmatrix},
    \begin{pmatrix}
        0 & 0 & 0\\
        0 & 0 & 0\\
        2\lambda & 0 & 0
    \end{pmatrix};
\end{equation}
and the self interaction $M_{\mathrm{self}}$:
\begin{equation}
    M_{\mathrm{self}}=
    \begin{pmatrix}
        0 & \lambda & \lambda\\
        \lambda & 0 & \lambda\\
        \lambda & \lambda & 0
    \end{pmatrix}.
\end{equation}

Let us define some important observables.
The magnetization vector $\vec{m}$ and its magnitude $m$ are defined by 
\begin{equation}
    \vec{m}:=\frac{1}{V}\sum_{n\in\Lambda}\vec{S}_{n},
\end{equation}
\begin{equation}
    m:=|\vec{m}|=\sqrt{\vec{m}\cdot\vec{m}}.
\end{equation}
The magnetic susceptibility $\chi_m$ and the (4th order) Binder parameter $U$ are given by~\cite{Binder:1981sa} 
\begin{equation}
    \chi_m:=V\left(\ev*{m^{2}}-\ev*{m}^{2}\right),
\end{equation}
\begin{equation}
    U:=1-\frac{\ev*{m^{4}}}{3\ev*{m^{2}}^{2}}.
\end{equation}
The value of the Binder parameter at the critical temperature $\beta_c = 1/T_c$ becomes $L$-independent in the large $L$ asymptotic region.
We use this property in the following analysis.

\section{Simulation details}
\label{sec_simulation_detail}
From now on, we fix $J=-1$ in the lattice Hamiltonian~\eqref{eq_lattice_model_calc} and consider the cubic lattice with $L$ and and $V = L^3$ under the periodic boundary condition.
The simulations are performed by changing $L$ from $6$ to $40$ in increments of $2$.
We take mainly $\lambda=0$, namely the original Heisenberg model, and $\lambda=8$ for the local Heisenberg-dipolar model, which is shown to be regarded near the dipolar fixed point in Section~\ref{subsec_validity_dipolar}. 

We focus on the narrow region of the temperature $\beta$ around the critical point.
For $\lambda = 0$, we select 20 points from the region $0.690608 \leq \beta \leq 0.695652$, 
and further, double the number of points when $L \geq 32$ by narrowing the interval.
For $\lambda = 8$, we similarly select 28 points from the region $0.319489 \leq \beta \leq 0.321012$, 
and double the number of points by extending the range to $3.11515 \leq \beta \leq 0.321012$ when $L \leq 20$ 
to include all the intersection points of the Binder parameter.

In the Monte Carlo simulation, we utilize the Julia package~\footnote{\url{https://github.com/fbuessen/SpinMC.jl}} of \texttt{SpinMC.jl} for the simulation.
We use the Metropolis algorithm with the replica exchange between different temperatures.
In each replica, the spin variables are randomly selected and updated one by one, 
and a series of updates equal to the lattice volume $V$ is called one sweep.
We evaluate the typical autocorrelation time $\tau$ of the magnetic susceptibility $m$ 
around the critical temperature for the largest lattice size $L=40$, 
where the autocorrelation becomes the longest in our simulations.
We find $\tau=O(100)$ and $O(1000)$ sweeps for $\lambda=0$ and $8$, respectively.
Since the autocorrelation becomes about 10 times longer due to the effect of the local dipolar constraint, 
we measure the observables every 100 sweeps for $\lambda=0$ and every 1000 sweeps for $\lambda=8$.
We also skip the first 1000 measurements for thermalization.

\subsection{Effectiveness of the local dipolar constraint}
\label{subsec_validity_dipolar}
Before examining the critical phenomena of the local Heisenberg-dipolar model, we would like to check how strictly the dipolar constraint is imposed for a different value of $\lambda$. As we will argue, taking the $\lambda \to \infty$ limit is extremely hard, so it is important to see the effectiveness of the local dipolar constraint for a finite value of $\lambda$. 

We measure the expectation values of the Heisenberg part, $A_n := \sum_i\ev*{\vec{S}_n \cdot \vec{S}_{n+\hat{i}}}$, 
and the local dipolar constraint term, $B_n := \ev*{(\vec{\nabla} \cdot \vec{S}_n)^2}$, of the Hamiltonian~\eqref{eq_lattice_model_calc}, 
using generated Monte Carlo samples.
Here, we take the lattice size $L=40$ 
and set the temperature to $\beta=0.69019$, $0.35288$, and $0.320374$ for $\lambda=0$, $4$, and $8$, respectively.
These values of $\beta$ are indeed the peak positions of the magnetic susceptibility $\chi_m$.
(See Appendix~\ref{sec_chi_peak} for the details of these choices.)
\begin{table}[h]
    \centering 
    \begin{tabular}{|c|c|c|}
        \hline
        $\lambda$ & $\sum_n A_n/V$ & $\sum_n B_n/V$ \tabularnewline
        \hline \hline
        0 & 0.9776(2) & (1.3486(3)) \tabularnewline
        \hline
        4 & 0.6880(3) & 0.30243(5) \tabularnewline
        \hline
        8 & 0.6521(3) & 0.17909(2) \tabularnewline
        \hline
    \end{tabular}
    \caption{\label{tab_A_and_B}
    The average values of $A_n$ and $B_n$ over the whole lattice sites are summarized for each $\lambda$, 
    which correspond to the first and second terms of the Hamiltonian~\eqref{eq_lattice_model_calc}. 
    Note that we write a bracket on $B_n$ at $\lambda=0$ since it does not contribute to the value of the Hamiltonian in simulations. }
\end{table}
The average values of them over the whole lattice sites are shown in Table~\ref{tab_A_and_B}.
We find that the local dipolar constraint term and the Heisenberg term give the same order of contribution 
to the Hamiltonian given by $\ev*{H}=\sum_n (-A_n + \lambda B_n)$.
This means that larger $\lambda$ gives the more strict imposition of the constraint as expected. Furthermore, the results for $\lambda = 4$ and $8$ are roughly consistent with the virial theorem. 

\begin{figure}[h]
    \centering
    \includegraphics[width=0.8\columnwidth]{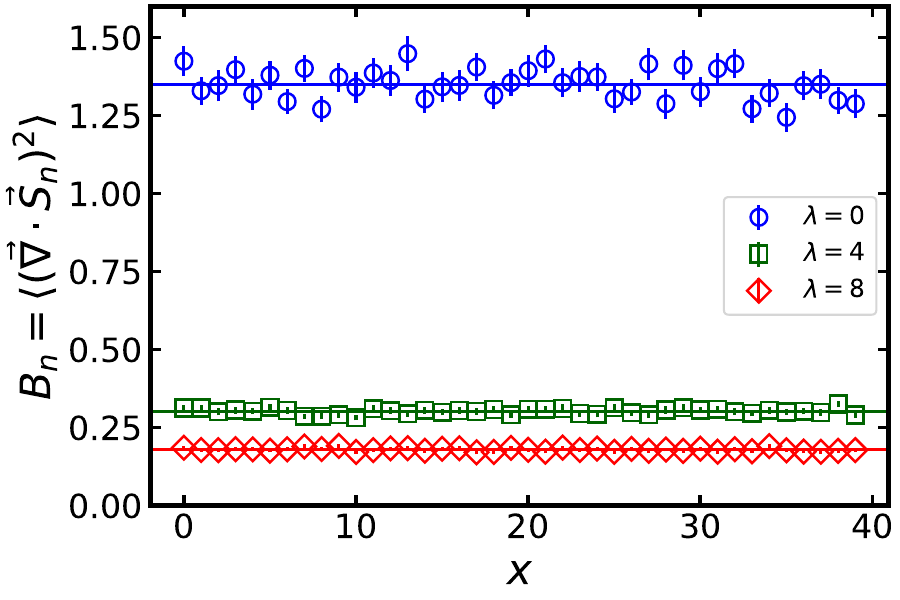}
    \caption{\label{fig_dipolar_const}
    The results of $B_n = \ev*{(\vec{\nabla} \cdot \vec{S}_n)^2}$ 
    are plotted against the coordinate $x$ on the $x$-axis $n=(x,0,0)$.
    The lattice size is set to $L=40$.
    The symbols of circle-blue, square-green, and diamond-red 
    correspond to $\lambda=0$, $4$, and $8$, respectively.
    The colored horizontal lines depict the average values over the whole lattice sites.}
\end{figure}

We further investigate a local property of the dipolar constraint.
In the Hamiltonian~\eqref{eq_lattice_model_calc}, the dipolar constraint term forces the spin $\vec{S}_n$ to satisfy $(\vec{\nabla} \cdot \vec{S}_n)^2=0$ at each site $n$.
Here, we measure $B_n$ locally at each site $n=(x,0,0)$ as depicted in Figure~\ref{fig_dipolar_const} for $\lambda=0$ (circle-blue symbol), $4$ (square-green), and $8$ (diamond-red).
As we can see in Figure~\ref{fig_dipolar_const}, 
the data points of $B_n$ for each site fluctuate around the average value and approach zero as $\lambda$ increases, 
which indicates that the dipolar constraint is imposed uniformly site-by-site.
Moreover, we will show that the shapes of the two-point correlation functions of the spin variables at $\lambda=4$ and $8$ are consistent with each other, while they show a clear difference from the one at $\lambda=0$. 
Consequently, we regard the local Heisenberg-dipolar model at $\lambda=8$ as a theory near the dipolar fixed point.

Ideally, we would like to take the $\lambda \to \infty$ limit. Unfortunately, within our Monte Carlo simulation method, taking the larger $\lambda$ shows us a technical challenge. The main reason is the longer autocorrelation time with increased $\lambda$ mentioned above, which hinders us from performing the Monte Carlo simulation efficiently. In this paper, we restrict ourselves to the simulation up to $\lambda = 8$.

\section{Determination of critical exponents}
\label{sec_result}

\subsection{Finite-size scaling and calculation strategy}\label{sec_scaling}
To evaluate critical exponents from Monte Carlo simulations, there are several different approaches: (1) to study power laws of susceptibility $\chi_m(\beta)$ and specific heat $C(\beta)$ as a function of $(\beta-\beta_c)$, (2) to study finite-size scaling of $U(\beta_c)$ and $\chi_m(\beta_c)$ at the critical temperature (3) to study correlation functions. Empirically, it is known that (2) gives the most precise (and hopefully most accurate) determination of critical exponents.
For instance, in Ref.~\cite{Hasenbusch:2020pwj}, the author gives one of the state-of-art results on the critical exponents from a lattice model, which is the same universality-class model with the Heisenberg model.  We essentially adopt this approach to obtain the exponents.

The finite-size scaling for the observables $U(\beta, L)$ and $\chi_m(\beta, L)$ is expressed as
\begin{equation}
    \left. \frac{\partial U(\beta, L)}{\partial\beta} \right|_{\beta=\beta_{c}} \propto L^{1/\nu},
    \label{eq_scaling_U}
\end{equation}
\begin{equation}
    \left. \chi_m(\beta, L) \right|_{\beta=\beta_{c}} \propto L^{2-\eta},
    \label{eq_scaling_chi}
\end{equation}
with these exponents $\nu$ and $\eta$.
Thus, once we obtain the value of the critical temperature $\beta_{c}$, 
we can estimate the critical exponents from these scaling relations.
Therefore, our first task is to determine the critical temperature $\beta_{c}$ in high precision.
Here, we use the property of the Binder parameter, which becomes independent of $L$ up to possible finite-size corrections.
Furthermore, to estimate $\beta_c$ more precisely, we take into account its finite-size corrections by following the steps below.

From the Monte Carlo simulation, we first measure the Binder parameter for several lattice sizes and $\beta$.
Given two different lattice sizes $L_1$ and $L_2$, the two curves of $U(\beta, L_1)$ and $U(\beta, L_2)$ 
intersect near the critical point $(\beta_c, U^*)$ on the $\beta\, \text{-}\, U$ plane.
We denote the intersecting positions as coordinates $(\beta_\times, U_\times)$ on the $\beta\, \text{-}\, U$ plane for each choice of $L_1$ and $L_2$.

Due to the finite-size effects, $(\beta_\times, U_\times)$ depends on $L_1$ and $L_2$, but we can estimate its dependence from an appropriate finite-size correction ansatz.
Here we consider a simple correction term of $O(L^{-\omega})$ with the exponent $\omega > 0$ 
and assume the asymptotic behavior $U(\beta_{c},L) \simeq c_{\omega}L^{-\omega}+\mathrm{const.}$.
Then, based on the scaling relation~\eqref{eq_scaling_U}, 
we have an ansatz on the behavior of the Binder parameter around $\beta_c$ as in Ref.~\cite{Binder:1981sa}, 
\begin{equation}
    U(\beta,L) = U^{*} + c_{\omega}L^{-\omega} + c_{\nu}L^{1/\nu}\frac{\beta-\beta_{c}}{\beta_{c}}.
    \label{eq_ansatz_U}
\end{equation}
This ansatz gives us an explicit prediction for the intersection point for finite $L_1$ and $L_2$ by solving an equation $U(\beta, L_1) = U(\beta, L_2)$. To present the solution more explicitly, here we introduce (recall $y_t = 1/\nu$)
\begin{equation}
\begin{aligned}
    f(L_1, L_2)&:=-\frac{L_{2}^{-\omega}-L_{1}^{-\omega}}{L_{2}^{y_t}-L_{1}^{y_t}}, \\
    g(L_1, L_2)&:=\frac{L_{2}^{y_t}L_{1}^{-\omega}-L_{1}^{y_t}L_{2}^{-\omega}}{L_{2}^{y_t}-L_{1}^{y_t}},
    \label{eq_f_and_g}
\end{aligned}
\end{equation}
and then the solution $\beta = \beta_\times$ is given by 
\begin{equation}
    \beta_\times = \beta_c\left(1 + \frac{c_{\omega}}{c_{\nu}} f(L_1, L_2) \right).
    \label{eq_sol_beta_x}
\end{equation}
Substituting the solution into $U(\beta, L_1)$, we have 
\begin{equation}
    U_\times = U^* + c_{\omega}\, g(L_1, L_2).
    \label{eq_sol_U_x}
\end{equation}

Once we determined the optimal values of the fitting parameters $\omega$ and $y_t$, we may be able to test the quality of the fitting by plotting measured $\beta_\times$ and $U_\times$ as a function of $f$ and $g$ (by changing $L_1$ and $L_2$ with substituting the optimal $\omega$ and $y_t$ into Eq.~\eqref{eq_f_and_g}). Our finite-size scaling ansatz predicts the linear dependence as shown in Figure~\ref{fig_bUx_fit} later. One merit of this linearity test is that we can see the quality of the fitting graphically rather than by reporting $\chi^2$.

Finally, let us summarize a calculation strategy to obtain the critical temperature ($\beta_c$) 
and several critical exponents ($\nu, \eta, \gamma$).
First, we obtain $\beta_c$ and $\nu$ from the data of the Binder parameter $U$.
\begin{description}
    \item[Step 1 for $\nu$] Compute the raw data of $U$ for various $\beta$ and $L$ using the Monte Carlo simulation
    \item[Step 2 for $\nu$] Fit the result of $U$ by the interpolating function $\tilde{U}(\beta) = c_0 + c_1\beta + c_2\beta^2$ for each $L$
    \item[Step 3 for $\nu$] Obtain the intersection points $(\beta_\times, U_\times)$ 
    using the obtained functions $\tilde{U}(\beta)$ for different lattice sizes
    \item[Step 4 for $\nu$] Fit the data points of $(\beta_\times, U_\times)$ by the ansatz~\eqref{eq_sol_beta_x} and~\eqref{eq_sol_U_x} 
    to determine $\beta_c$ with high precision
    \item[Step 5 for $\nu$] Compute the gradient $\partial \tilde{U}(\beta)/\partial\beta = c_1 + 2c_2\beta$ at $\beta=\beta_c$ for each $L$ 
    and fit them based on Eq.~\eqref{eq_scaling_U} to obtain the exponent $\nu$
\end{description}
Next, we perform a similar calculation for the magnetic susceptibility $\chi_m$ to obtain the critical exponent $\eta$:
\begin{description}
    \item[Step 1 for $\eta$:] Compute the raw data $\chi_m$ for various $\beta$ and $L$ using the Monte Carlo simulation
    \item[Step 2 for $\eta$:] Fit the result of $\chi_m$ by the interpolating function 
    $\tilde{\chi}_m(\beta) = d_0 + d_1\beta + d_2\beta^2 + d_3\beta^3$ for each $L$
    \item[Step 3 for $\eta$:] Compute $\tilde{\chi}_m(\beta)$ at $\beta=\beta_c$ for each $L$ 
    and fit them based on Eq.~\eqref{eq_scaling_chi} to obtain the exponent $\eta$
\end{description}
Then, we obtain the exponent $\gamma$ using the relation $\gamma=\nu(2-\eta)$.

Taking into account the autocorrelation time mentioned in Section~\ref{sec_simulation_detail}, we measured observables with 100 and 1000 sweeps intervals for $\lambda=0$ and $8$, respectively. In other words, each measurement data can be regarded as almost independent.
We perform more than $10^5$ measurements for each parameter set of ($\beta, L, \lambda$).
To estimate statistical errors, the entire analysis of the procedures listed above is performed using the jackknife method with the binned data, where the measurement data for each simulation parameter is divided into 100 bins. In this way, autocorrelation and error propagation were fully taken into account in the evaluation of statistical errors.

\subsection{Intersecting points of the Binder parameter}
\label{subsec_intersecting_binder}
\begin{figure*}[tb]
    \centering
    \includegraphics[width=0.7\columnwidth]{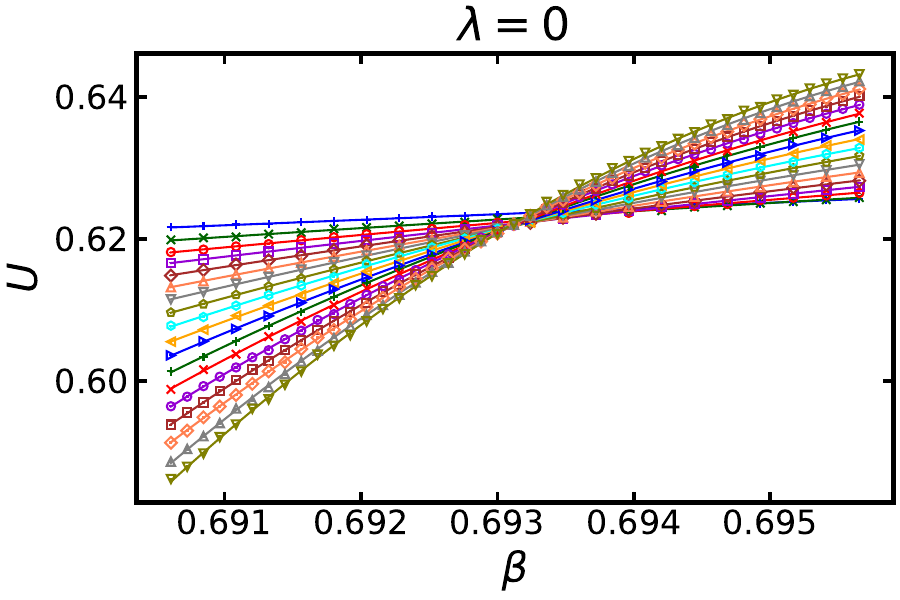}\ 
    \includegraphics[width=0.7\columnwidth]{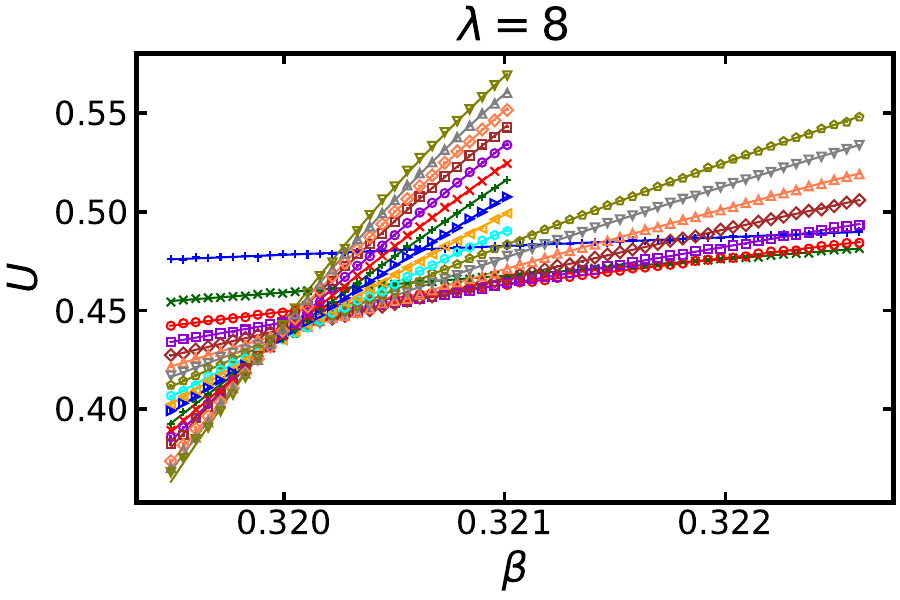}
    \includegraphics[width=0.28\columnwidth]{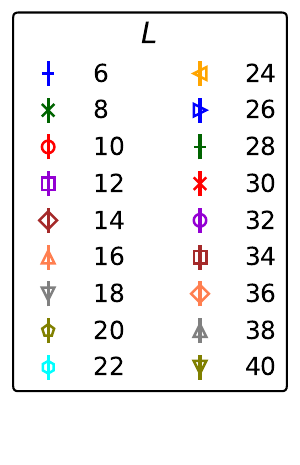}
    \caption{\label{fig_raw_U}
    The raw data of the Binder parameter $U$ are plotted against the temperature $\beta$.
    The left and right panels correspond to the results of the Heisenberg model $\lambda=0$ 
    and the local Heisenberg-dipolar model $\lambda=8$, respectively.
    The fitting results by the interpolating function $\tilde{U}(\beta)$ are also shown for each $L$.}
\end{figure*}

The raw data of the Binder parameter $U$ is shown in Figure~\ref{fig_raw_U}, 
where we focus on the narrow range of $\beta$ around the critical temperature $\beta_c$.
The results of $U$ for the different system sizes $L$ intersect at approximately the same point 
because of the asymptotic volume independence at $\beta_c$. As explained in the previous subsection, the deviation of the intersection point is caused by the finite-size effect, 
which becomes more significant as $L$ decreases.
To take care of this effect, we first compute $\beta_c$ using the finite-size scaling methods for the intersection points.

To find the intersection point precisely, 
we introduce the interpolating function of $\beta$ for the discrete data points of $U$, 
\begin{equation}
    \tilde{U}(\beta)|_L := c_0(L) + c_1(L) \beta + c_2(L) \beta^2,
    \label{eq_interpolate_U}
\end{equation}
where the parameters $\{c_i(L)\}$ are determined by fitting for each $L$ individually.
The fitting results are shown in Figure~\ref{fig_raw_U} as the solid curves.
See also Table~\ref{tab_c_and_d} in Appendix~\ref{sec_fit_result_interpolate}, where the best-fit values of $\{c_i\}$ are presented.

\begin{figure*}[tb]
    \centering
    \includegraphics[width=0.8\columnwidth]{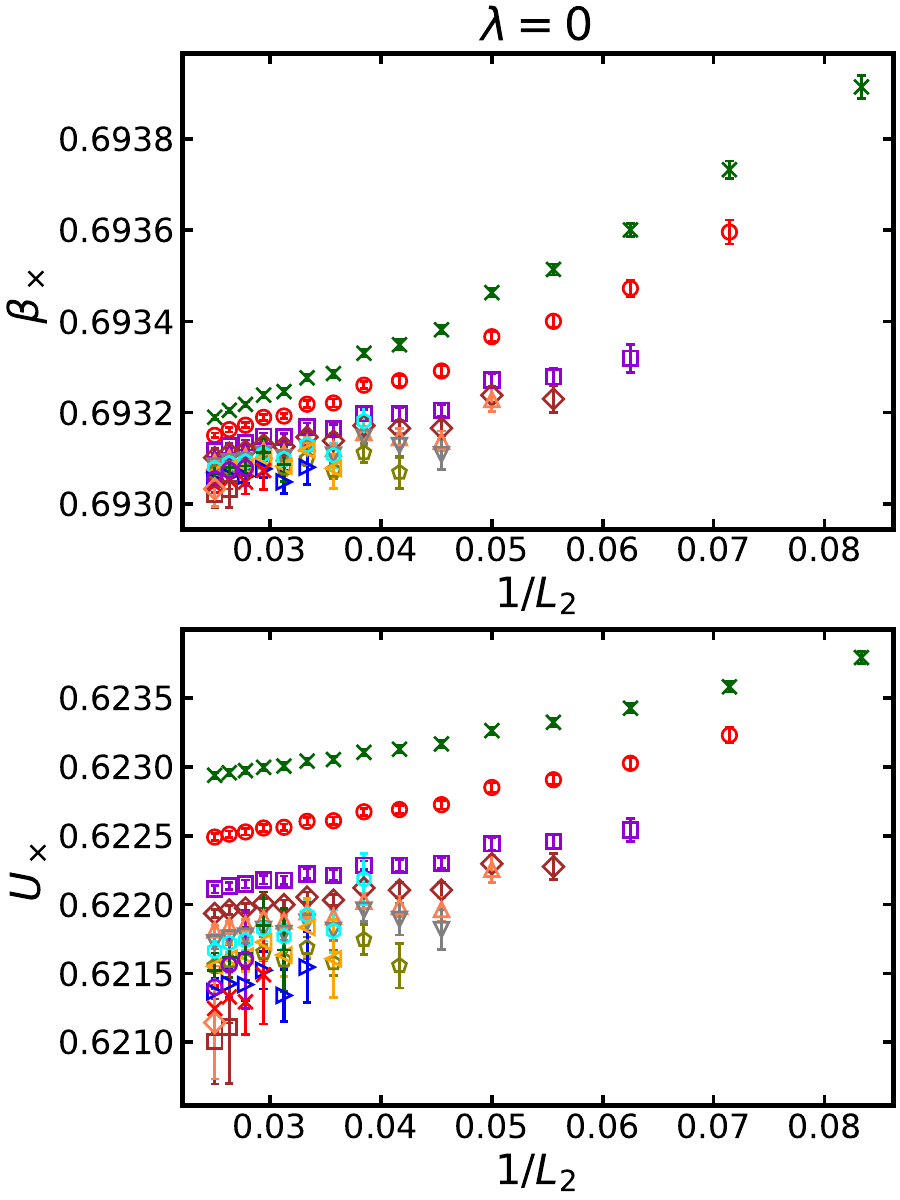}
    \quad
    \includegraphics[width=0.8\columnwidth]{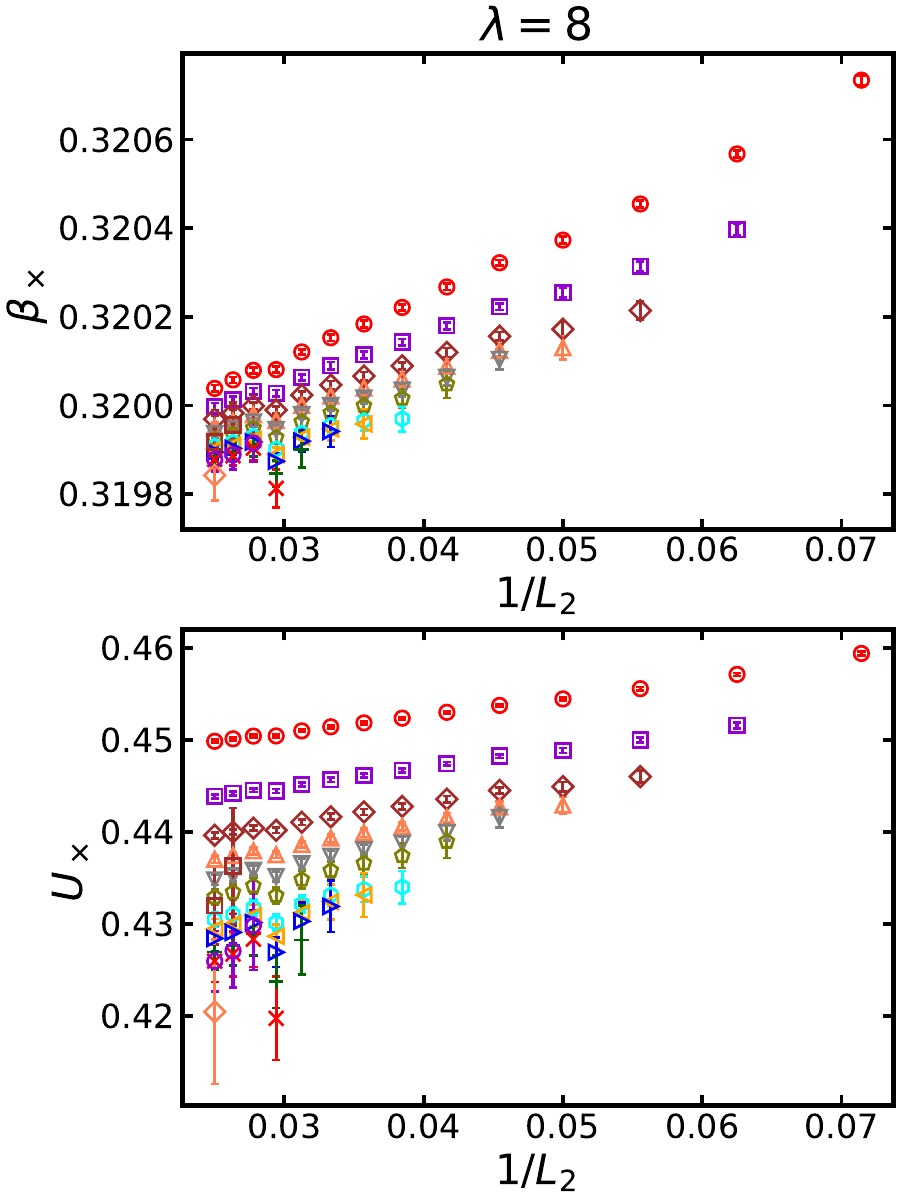}
    \caption{\label{fig_Ux_vs_bx}
    The intersecting positions $\beta_\times$ (top) and $U_\times$ (bottom) of the Binder parameter 
    for the two lattice sizes $L_1$ and $L_2$ with $L_1 < L_2$ are plotted against $1/L_2$ for $\lambda=0$ (left) and $8$ (right).
    The data points with the same $L_1$ are denoted by the same symbols as $L$ shown in the legend of Figure~\ref{fig_raw_U}.
    The data points converge to the critical values, $\beta_c$ and $U^*$, as both $L_1$ and $L_2$ increase.}
\end{figure*}

Now, we can calculate the intersection points $(\beta_\times, U_\times)$, 
using the interpolating function~\eqref{eq_interpolate_U} for two different lattice sizes $L_1$ and $L_2$.
First, we solve the quadratic equation, $\tilde{U}(\beta)|_{L_1} = \tilde{U}(\beta)|_{L_2}$, and obtain $\beta_\times$ as a solution.
Then, substituting the value of $\beta_\times$ into $\tilde{U}(\beta)|_{L_1}$, we obtain the value of $U_\times$ as well.
The results of the intersection points, $\beta_\times$ and $U_\times$, for $L_1 < L_2$ 
are plotted against $1/L_2$ in Figure~\ref{fig_Ux_vs_bx}.
The data points corresponding to the same $L_1$ are denoted by the same symbols as $L$ in Figure~\ref{fig_raw_U}.

In Figure~\ref{fig_Ux_vs_bx}, the intersection points are not gathered in one place but somewhat spread out depending on $L_1$ and $L_2$.
This suggests a nonnegligible finite-size correction as we expected in the ansatz~\eqref{eq_ansatz_U}, namely $c_{\omega} \neq 0$.
Since the distributions of the points for $\lambda=0$ and $8$ are qualitatively the same, 
the common finite-size scaling ansatz should apply to both cases.
As $L_1$ and $L_2$ increase, the data points approach 
$(\beta, U)\approx(0.6930, 0.621)$ for $\lambda=0$ and $(\beta, U)\approx(0.3198, 0.41)$ for $\lambda=8$, 
where the critical points $(\beta_c, U^*)$ seem to be located around.

\subsection{Determination of the critical temperature}
\label{subsec_determine_bc}
Now, we determine the critical temperature $\beta_c$ by using the scaling property of intersections of the Binder parameter $U$ since the explicit value of $\beta_c$ is necessary to compute the critical exponents.
We perform the simultaneous fitting of $(\beta_\times, U_\times)$ by using Eqs.~\eqref{eq_sol_beta_x} and~\eqref{eq_sol_U_x} to find optimal parameters $(\beta_c, U^*, y_t, \omega, c_{\nu}, c_{\omega})$.

\begin{table*}[tb]
    \centering 
    \begin{tabular}{|c|c|c|c|c|c|c|c|c|}
        \hline
        $\lambda$ & $L_{\mathrm{min}}$ & $\beta_c$ & $U^*$ & $y_t$ & $\omega$ & $c_\nu$ & $c_\omega$ & $\chi^2/\mathrm{dof}$ \tabularnewline
        \hline \hline
          & 6 & 0.693015(8) & 0.6206(1) & 1.42(1) & 1.1(1) & 0.020(2) & 0.043(1) & 1.51 \tabularnewline
        \cline{2-9}
        0 & 8 & 0.693035(11) & 0.6210(2) & 1.42(1) & 1.4(2) & 0.035(1) & 0.043(1) & 1.22 \tabularnewline
        \cline{2-9}
          & 10 & 0.693054(13) & 0.6213(2) & 1.39(1) & 2.1(4) & 0.128(101) & 0.047(2) & 1.03 \tabularnewline
        \hline \hline
          & 8 & 0.319887(9) & 0.417(2) & 1.85(1) & 1.2(1) & 0.52(5) & 0.061(2) & 4.11 \tabularnewline
        \cline{2-9}
        8 & 10 & 0.319844(18) & 0.407(7) & 1.78(1) & 0.9(2) & 0.31(6) & 0.076(2) & 1.39 \tabularnewline
        \cline{2-9}
          & 12 & 0.319803(35) & 0.378(39) & 1.74(1) & 0.5(3) & 0.20(2) & 0.084(3) & 0.97 \tabularnewline
        \hline
    \end{tabular}
    \caption{\label{tab_fit_intersection}
    The fitting results of the intersection points $(\beta_\times, U_\times)$ with the different choices of $L_{\mathrm{min}}$ for $\lambda=0$ and $8$.
    In these results, $L_{\mathrm{diff}}=4$ and $L_{\mathrm{max}}=40$ are fixed.}
\end{table*}
In this analysis, we introduce $L_{\mathrm{diff}}$, $L_{\mathrm{min}}$, and $L_{\mathrm{max}}$ to control the fitting range as 
$L_2 - L_1 \ge L_{\mathrm{diff}}$ and $L_{\mathrm{min}} \leq L_1 < L_2 \leq L_{\mathrm{max}}$.
The fitting results are summarized in Table~\ref{tab_fit_intersection}, where we consider three different choices of $L_{\mathrm{min}}$ for each $\lambda$.
We fix $L_{\mathrm{diff}}=4$ to remove the closest pairs of $L_1$ and $L_2$ from the fitting.
This prescription excludes the situation in which the slopes of the functions $\tilde{U}(\beta)|_{L_1}$ and $\tilde{U}(\beta)|_{L_2}$ are almost the same, 
so that the noisy data of the intersection point are avoided.
Furthermore, when we choose $L_{\mathrm{min}}=8$ and $10$ for $\lambda=0$ and $8$, respectively, 
the resulting $\beta_c$ is almost independent of $L_{\mathrm{max}}$ up to the statistical error.
Thus, we determine the best-fit values of $\beta_c$ with these choices of $L_{\mathrm{min}}$ and $L_{\mathrm{max}}=40$, 
where the statistical errors are under control and $\chi^2/\mathrm{dof} \sim 1$.

\begin{figure*}[tb]
    \centering
    \includegraphics[width=0.8\columnwidth]{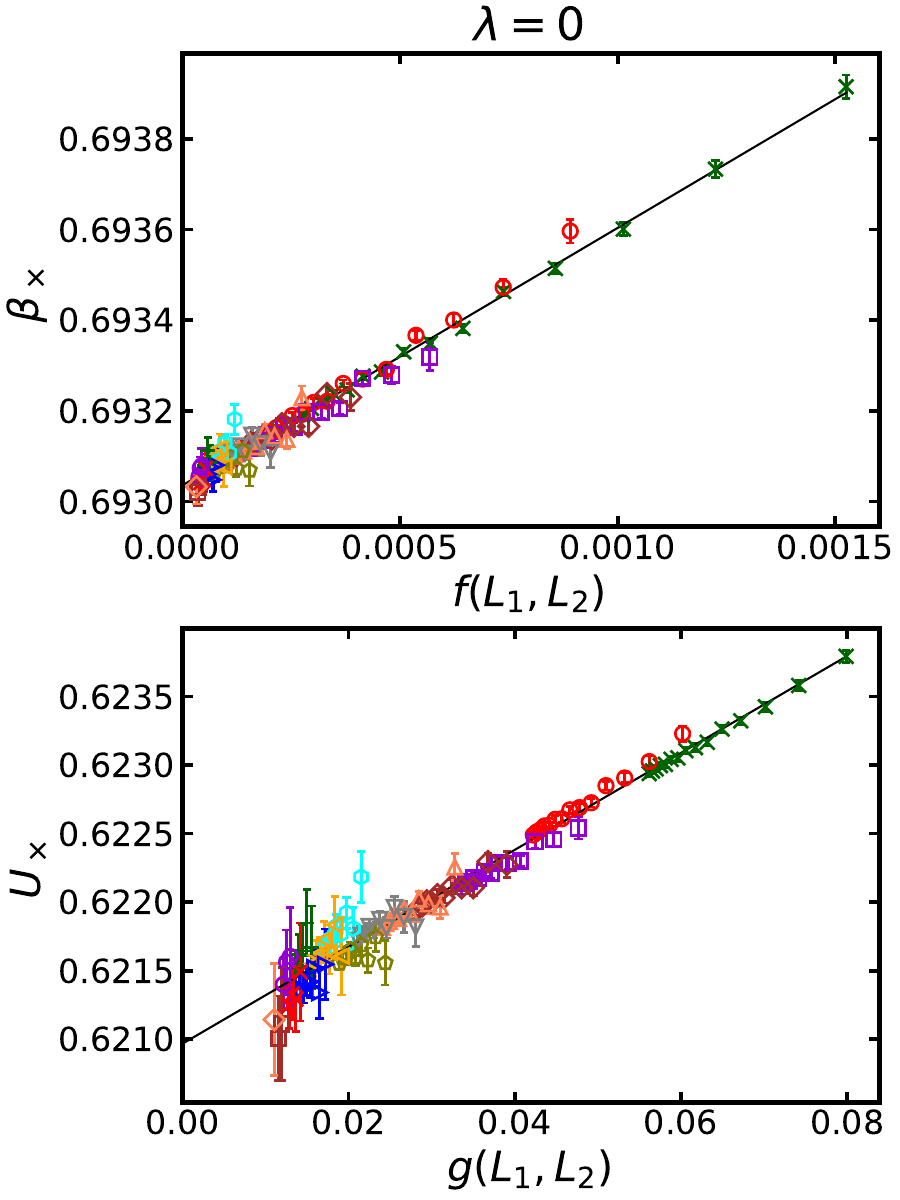}
    \quad
    \includegraphics[width=0.8\columnwidth]{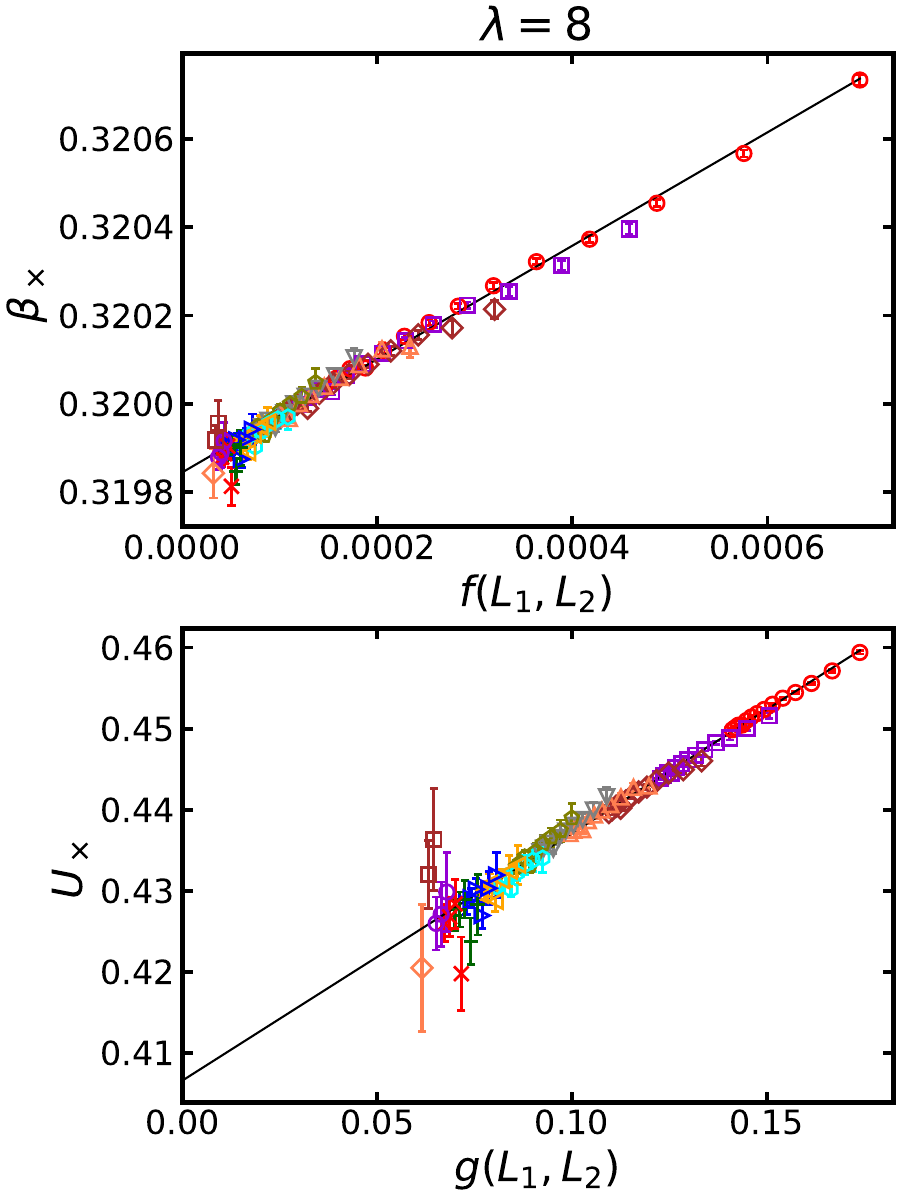}
    \caption{\label{fig_bUx_fit}
    The fitting results of the scaling behavior of the intersections $(\beta_{\times}, U_{\times})$ 
    are shown for $\lambda=0$ (left) and $8$ (right).
    The numerical results of $\beta_{\times}$ (top) and $U_{\times}$ (bottom) 
    are plotted against the functions $f(L_1, L_2)$ and $g(L_1, L_2)$, respectively, 
    where the best-fit values of $y_t$ and $\omega$ are used to define these functions.
    Note that the data points are aligned on straight lines, which indicates the validity of the fitting ansatz. Here, the data symbols denote the same ones by replacing $L$ to $L_1$ shown in the legend of Figure~\ref{fig_raw_U}.}
\end{figure*}
In Figure~\ref{fig_bUx_fit}, $\beta_\times$ and $U_\times$ are plotted against $f(L_1, L_2)$ and $g(L_1, L_2)$ defined in Eq.~\eqref{eq_f_and_g}, respectively, so that the linearity should indicate our finite-size scaling ansatz is working well.
Here, to compute $f(L_1, L_2)$ and $g(L_1, L_2)$, we use the best-fit values of $y_t$ and $\omega$, obtained with $L_{\mathrm{min}}=8$ ($10$) for $\lambda=0$ ($8$), in Table~\ref{tab_fit_intersection}.
The numerical data seen in Figure~\ref{fig_bUx_fit} shows visible linearity for both $\beta_\times$ and $U_\times$ with $\lambda = 0$ and $8$, 
suggesting the validity of our ansatz.

\begin{figure}[h]
    \centering
    \includegraphics[width=0.8\columnwidth]{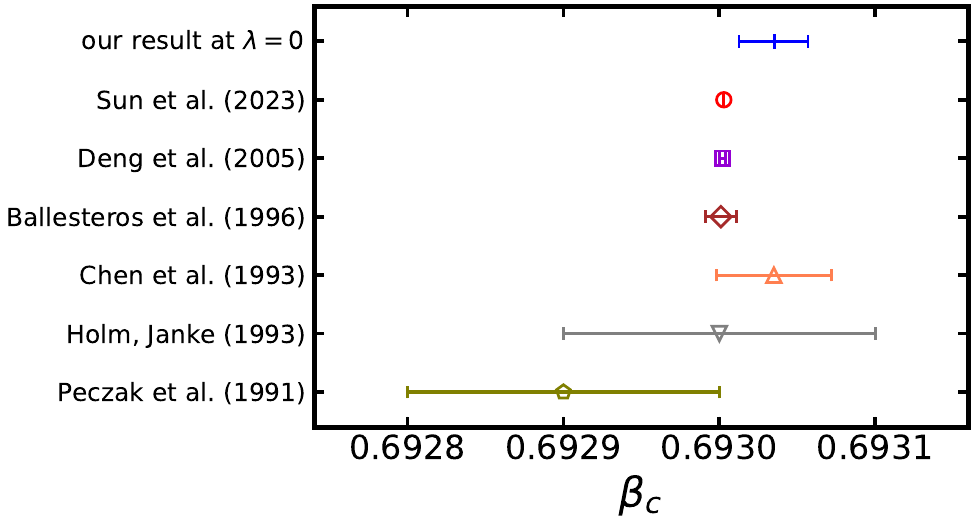}
    \caption{\label{fig_bc_comp}
    Comparison of the critical temperature $\beta_c$ of the Heisenberg model obtained in this paper and the previous studies 
    by the Monte Carlo~\cite{Sun:2023vwy, PhysRevE.72.016128, Ballesteros:1996bd, Chen:1993zz, Holm:1993zz, Peczak:1991zz}.}
\end{figure}
Given the success of the finite-size scaling ansatz, we obtain the critical temperatures for the two cases of $\lambda$ from the best-fit results as 
\begin{equation}
    \beta_c = 
    \begin{cases}
        0.693035(11)\left(^{+18}_{-20}\right) & (\lambda=0), \\
        0.319844(18)\left(^{+43}_{-41}\right) & (\lambda=8).
    \end{cases}
    \label{eq_beta_c}
\end{equation}
Here, the first parentheses denote the statistical error, which is given by the variance of the best-fit values for the 100 Jackknife samples.
The second parentheses represent the systematic error from the uncertainty of the finite-size effect, 
which is estimated by the differences in $\beta_c$ for changing $L_{\mathrm{min}} \to L_{\mathrm{min}} \pm 2$, 
namely the values in Table~\ref{tab_fit_intersection}.
We use these values of $\beta_c$ in the following analyses.
Our result of $\beta_c$ for $\lambda=0$ is compared with the previous Monte Carlo studies 
on the Heisenberg model in Figure~\ref{fig_bc_comp}.

\subsection{Critical exponent: \texorpdfstring{$\nu$}{nu}}
While we have already determined the values of $\nu = y_t^{-1}$ from the finite-size scaling ansatz~\eqref{eq_ansatz_U} (see Table~\ref{tab_fit_intersection}), 
there is a cleaner and more precise way to predict the critical exponent $\nu$. 
The idea is to study the $\beta$ derivative of Eq.~\eqref{eq_ansatz_U}, or equivalently the scaling ansatz~\eqref{eq_scaling_U}.
This ansatz is cleaner because we have removed three unknowns $U_*$, $c_\omega$, and $\omega$ which had contaminated the determination of $y_t$, 
so with the ansatz~\eqref{eq_ansatz_U} we expect the more precise determination of $\nu$ from the data is possible.

\begin{figure*}[tb]
    \centering
    \includegraphics[width=0.8\columnwidth]{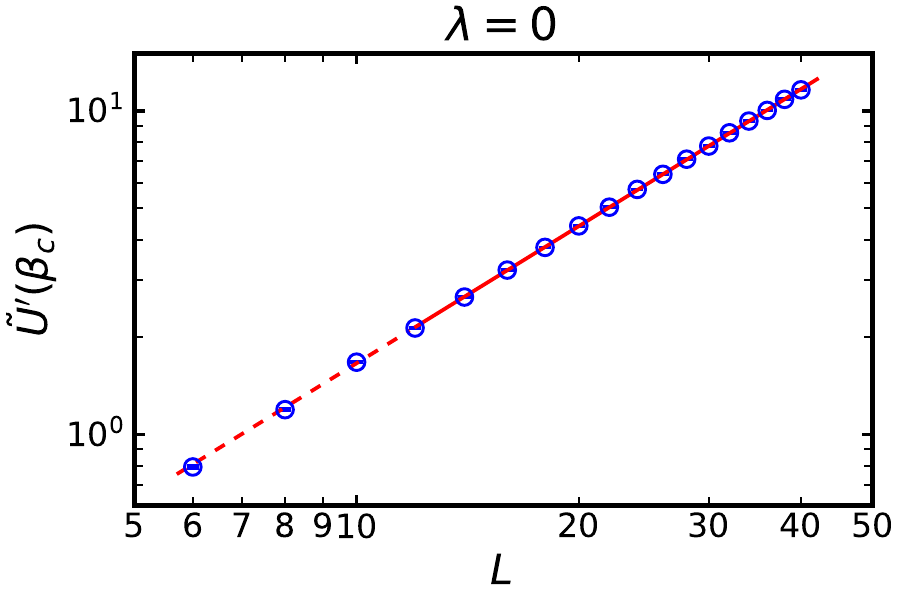}
    \quad
    \includegraphics[width=0.8\columnwidth]{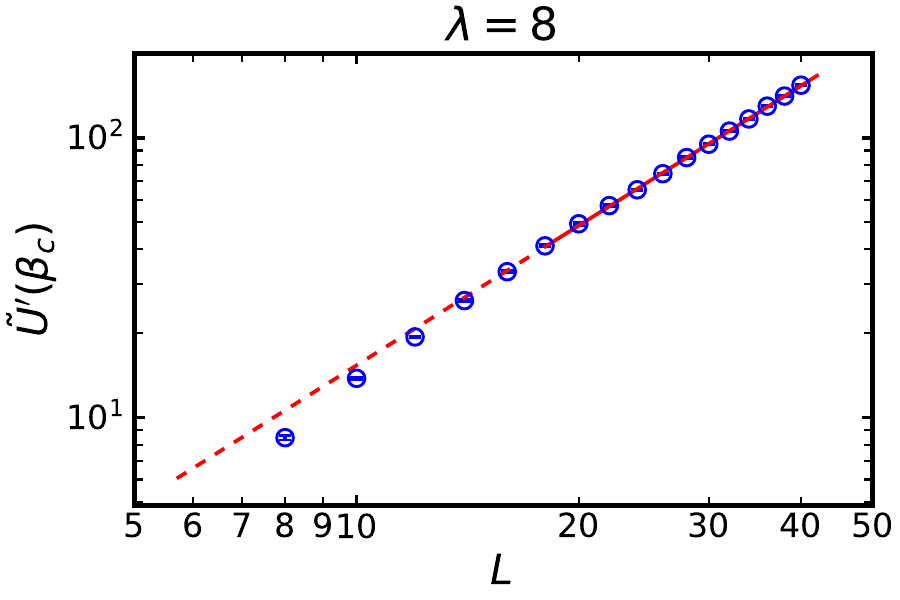}
    \caption{\label{fig_grad_vs_L} 
    The $L$ dependence of the gradient of the Binder parameter at the critical temperature $\beta_c$ is plotted in log-log scale for $\lambda=0$ (left) and $8$ (right).
    The data is obtained using the interpolating function~\eqref{eq_interpolate_chi} as $\tilde{U}^{\prime}(\beta_c)$. 
    The fitting results by $a_0 L^{1/\nu}$ are also depicted by the solid (dashed) lines inside (outside) the fitting range, which is $L \in [12, 40]$ and $[18, 40]$ for $\lambda=0$ and $8$, respectively.}
\end{figure*}

First we compute the gradient of $U$ from the interpolating function~\eqref{eq_interpolate_U} for each $L$,
\begin{equation}
    \tilde{U}^{\prime}(\beta)|_L := \frac{\partial \tilde{U}(\beta)|_L}{\partial\beta} = c_1(L) + 2c_2(L) \beta
\end{equation}
with the best-fit values of $\{c_i(L)\} $.
Substituting the value of critical temperature $\beta_c$, we obtain $\tilde{U}^{\prime}(\beta_c)$ as a function of $L$.
The results are shown in Figure~\ref{fig_grad_vs_L} in the log-log scale, which grows almost as a power of $L$.

To compare the data with the expected scaling behavior~\eqref{eq_scaling_U}, 
we fit the data points with a function $a_0 L^{1/\nu}$ of $L$ as shown in Figure~\ref{fig_grad_vs_L}.
We obtain the best-fit values $(\nu, a_0) = (0.711(1), 0.0652(4))$ and $(0.603(2), 0.336(7))$ 
with the fitting ranges $L \in [12, 40]$ and $[18, 40]$, for $\lambda=0$ and $8$, respectively.
These ranges are taken from typical choices in the detailed analysis below.

The slight deviation from the linear growth in Figure~\ref{fig_grad_vs_L} suggests that the inclusion of the finite size correction 
would give a more precise estimate of the critical exponent $\nu$.
To test the effect of the finite-size correction, 
we fit the data of $\tilde{U}^{\prime}(\beta_c)$ by the scaling function, 
\begin{equation}
    U^{\prime}_c(L) = a_0 L^{1/\nu} (1 + a_1 L^{-\epsilon}),
    \label{eq_fit_func_nu}
\end{equation}
based on the scaling property in Eq.~\eqref{eq_scaling_U}.
Here we include the term of $L^{-\epsilon}$ as a finite-size correction, which should appear when $L$ is small, 
following a similar analysis in Ref.~\cite{Campostrini:2000iw, 2019PhRvB.100v4517H, Hasenbusch:2020pwj}. 
In principle, the correction exponent $\epsilon$ should be related to the other critical exponents such as  $\omega$ (see Ref.~\cite{2019PhRvB.100v4517H}), 
but the precise determination of $\epsilon$ from the fits requires more precisions of data in larger $L$ asymptotics.
In this paper, as did in Refs.~\cite{Campostrini:2000iw, 2019PhRvB.100v4517H, Hasenbusch:2020pwj}, 
we fix the value of $\epsilon$ beforehand and estimate the optimal value of $\epsilon$ by the stability of the resulting fit.

After benchmark of several possibile choices of $\epsilon$, here we present the results for $\epsilon=2.0$, $3.0$, $6.0$, and $\infty$. 
The value $\epsilon = \infty$ corresponds to no corrections to the leading-order scaling ansatz, 
which does not suffer from the lack of precision in the large $L$ asymptotics.
With each choice of $\epsilon$, the fitting parameters are now $\nu$, $a_0$, and $a_1$, 
and we specify the fitting range as $L_{\mathrm{min}} \leq L$ by changing $L_{\mathrm{min}}$.

\begin{figure*}[tb]
    \centering
    \includegraphics[width=0.8\columnwidth]{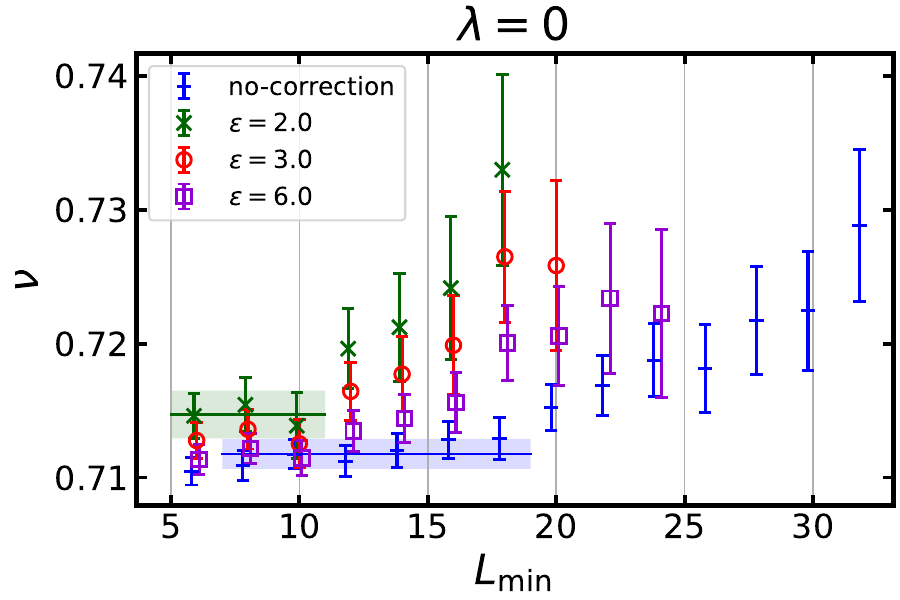}
    \quad
    \includegraphics[width=0.8\columnwidth]{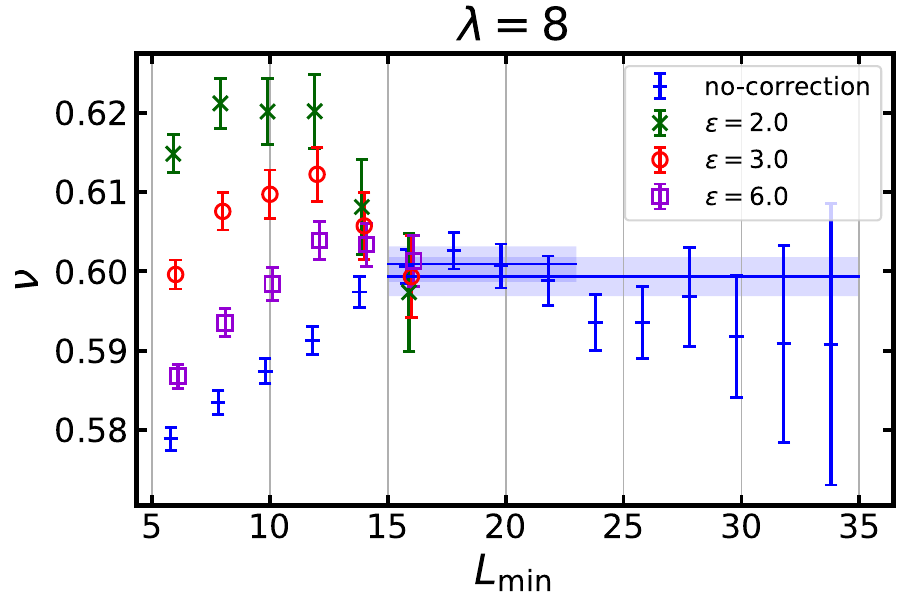}
    \caption{\label{fig_nu_vs_Lmin} 
    The results of $\nu$ obtained via the fitting ansatz~\eqref{eq_fit_func_nu} are plotted 
    against the lower bound $L_{\mathrm{min}}$ of the fitting range for $\lambda=0$ (left) and $8$ (right).
    In both panels, the symbols of plus-blue, cross-green, circle-red, and square-purple depict 
    the results with $\epsilon=\infty$, $2.0$, $3.0$, and $6.0$, respectively.
    The horizontal lines depict the results of the constant fitting with the error bands.}
\end{figure*}
In Figure~\ref{fig_nu_vs_Lmin}, the fitting results of $\nu$ with $\epsilon = 2.0$, $3.0$, $6.0$, and $\infty$ 
are plotted against $L_{\mathrm{min}}$ for $\lambda = 0$ and $8$.
For $\lambda = 0$ in the left panel, the data points of $\nu$ with the corrections, 
$\epsilon=2.0$ (cross-green symbol), $3.0$ (circle-red), and $6.0$ (square-purple), 
show plateau behavior in a narrow region $L_{\mathrm{min}} \lesssim 10$.
On the other hand, the ones of $\nu$ for the no-correction ansatz $\epsilon=\infty$ (plus-blue) 
are almost constant in a wider region $L_{\mathrm{min}} \lesssim 18$.
Furthermore, the plateau values for all $\epsilon$ agree with each other within $2\sigma$ error bar.
If $L_{\mathrm{min}}$ is too large, the data points have large errors and deviate from the plateau due to over-fitting.

Therefore, we perform the constant fitting of the data points in these plateau regions 
and depict the results as shadow bands in Figure~\ref{fig_nu_vs_Lmin}. 
As a central value, we pick up the result of the no-correction ansatz ($\epsilon=\infty$), then it results in $\nu = 0.712(1)$. 
Here, the statistical error is obtained by the fitting error.
To evaluate the systematic error from the ambiguity of $\epsilon$, 
we take the difference of the constant-fitting results for the other values of $\epsilon$ 
and obtain $\nu = 0.715(2)$ in the case of $\epsilon=2.0$ as the largest deviation.

Combining these results, the final result of the exponent $\nu$ is given by 
\begin{equation}
    \nu = 0.712(1) \left(^{+3}_{-0}\right) \left(^{+1}_{-1}\right)_{\beta_c} \qquad \text{for } \lambda = 0
\end{equation}
for the Heisenberg model. 
Here, the first and second parentheses denote the statistical error by the Jackknife method and the systematic error from the uncertainty of $\varepsilon$, respectively.
The third error comes from the propagation of the systematic error of $\beta_c$.

The situation is different for $\lambda=8$ as shown in the right panel of Figure~\ref{fig_nu_vs_Lmin}, 
where plateaus do not appear in the small $L_{\mathrm{min}}$ region even if $\epsilon$ is changed.
Instead, the result of the no-correction ansatz ($\epsilon=\infty$) becomes almost plateau when $L_{\mathrm{min}} \geq 16$.
If we introduce the finite $\epsilon$ term, the fluctuation of the fitting is sizable, and we cannot find a plateau regime. 
It suggests that the no-correction ansatz is enough to fit the data, and adding the correction term causes the over-fitting.
Indeed, the data points for any $\epsilon$ here coincide at $L_{\mathrm{min}} = 16$, 
where the plateau of the no-correction ansatz starts.

Compared with the case of $\lambda=0$, the scaling region (plateau) for $\lambda=8$ shifts toward larger $L_{\mathrm{min}}$, 
which should be caused by the dipolar constraint term.
The non-plateau behavior for smaller $L_{\mathrm{min}}$ implies that the ansatz~\eqref{eq_fit_func_nu} with the single correction term of $L^{-\epsilon}$ does not cover the finite size effect well.
Unfortunately, adding more correction terms to the ansatz is not feasible in the current precision of the data.
Thus, we use only the no-correction ansatz with the range $L_{\mathrm{min}} \geq 16$ in this analysis.
As a central result for $\lambda=8$, we perform the constant fitting over the result of the no-correction ansatz in the range $L_{\mathrm{min}} \in [16, 22]$, and then obtain $\nu = 0.601(2)$.
We also fit the same result in a wider region $L_{\mathrm{min}} \in [16, 34]$ 
to evaluate the systematic error from the ambiguity of the fitting range, resulting in $\nu = 0.599(2)$.

Finally, we obtain the result of the exponent as 
\begin{equation}
    \nu = 0.601(2) \left(^{+0}_{-2}\right) \left(^{+5}_{-4}\right)_{\beta_c} \qquad \text{for } \lambda = 8
\end{equation}
for the local Heisenberg-dipolar model.

These results of $\nu$ above are compared with the results of previous studies with various methods as shown in Figure~\ref{fig_nu_comp}.
Here, we compare the values of $\nu$ for $\lambda=0$, $8$, and $\infty$ obtained by the Monte Carlo~\cite{
Hasenbusch:2020pwj, Hasenbusch:2011zwv, Hasenbusch:2000ph, 
Ballesteros:1996bd, Chen:1993zz, Holm:1993zz, Peczak:1991zz, Lau:1989zz}, 
the conformal bootstrap~\cite{Chester:2020iyt}, 
the $\varepsilon$-expansion~\cite{Aharony_1, 2022NuPhB.98515990K}, 
and the functional renormalization group~\cite{Nakayama:2023wrx}.
For reference, the conformal bootstrap predicts $\nu=0.7117(2)$~\footnote{
Here, the error in the bootstrap bound is not one-sigma range but rigorous~\cite{Chester:2020iyt}.}
and the latest Monte Carlo study~\cite{Hasenbusch:2020pwj} reports $\nu=0.71164(10)$ for the Heisenberg model.
Our result at $\lambda=0$ is consistent with them. 
On the other hand, the result at $\lambda=8$ shows a large deviation from the analytical prediction by the 1-loop $\varepsilon$-expansion.
Here, we note that the difference between the 1-loop result of the $\varepsilon$-expansion and the 3-loop result at $\lambda=0$ is larger than the one between the 1-loop one at $\lambda=0$ and $\lambda=8$.
It indicates that loop corrections are sizable, at least larger than the $\lambda$-dependence.
    
Furthermore, let us discuss the self-consistency of our results.
As shown in Table~\ref{tab_fit_intersection}, 
we obtained the estimations of $y_t = 1/\nu$ by analysing the intersections of the Binder parameter in Section~\ref{subsec_determine_bc}, 
which yields $\nu=0.706(5)$ for $\lambda=0$ with $L_{\mathrm{min}}=8$ and $\nu=0.563(4)$ for $\lambda=0$ with $L_{\mathrm{min}}=10$.
Compared with these values, the results of $\nu$ obtained in this section have smaller statistical errors.
The smaller statistical error is mainly because we have fewer parameters to fit by considering the derivative of the Binder parameter and fixing $\beta_c$.
The results of the two methods agree within the error for $\lambda=0$ but not well for $\lambda=8$. 
One of the reasons for this discrepancy could arise mainly from the choice of $L_{\mathrm{min}}$, 
where we set $L_{\mathrm{min}}=10$ in Section~\ref{subsec_determine_bc} and $L_{\mathrm{min}} \ge 16$ in this section.

\subsection{Critical exponents: \texorpdfstring{$\eta$ and $\gamma$}{eta and gamma}}
\label{subsec_exponent_gamma}
Next, we move on to the critical exponent $\eta$.
We compute it via the scaling relation~\eqref{eq_scaling_chi} of the magnetic susceptibility $\chi_m$.
We fit the raw data of $\chi_m$ using the interpolating function, 
\begin{equation}
    \tilde{\chi}_m(\beta)|_L = d_0(L) + d_1(L) \beta + d_2(L) \beta^2 + d_3(L) \beta^3,
    \label{eq_interpolate_chi}
\end{equation}
for each $L$ individually as shown in Figure~\ref{fig_raw_chi}.
See also Table~\ref{tab_c_and_d} in Appendix~\ref{sec_fit_result_interpolate} for the best-fit values of $\{d_i\}$.
\begin{figure*}[tb]
    \centering
    \includegraphics[width=0.8\columnwidth]{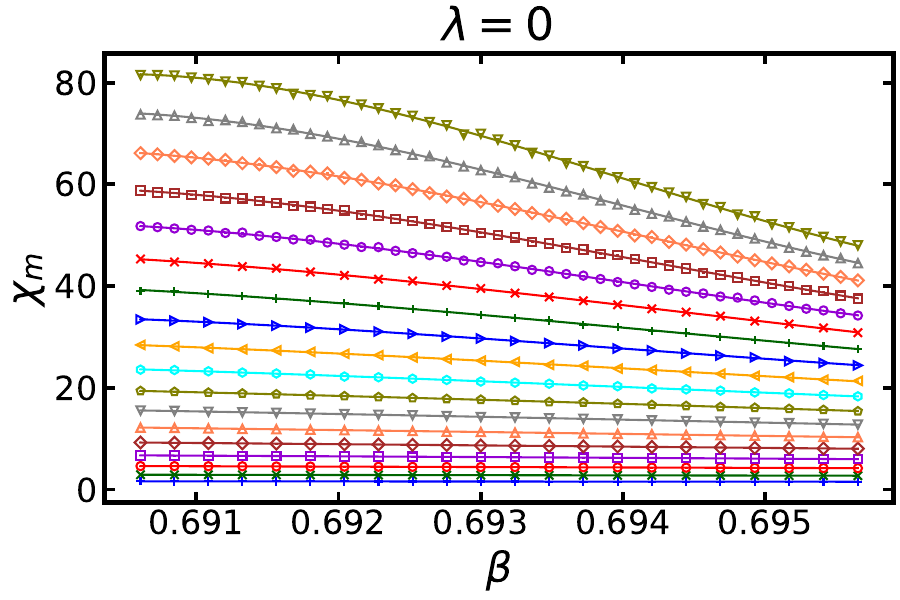}
    \quad
    \includegraphics[width=0.8\columnwidth]{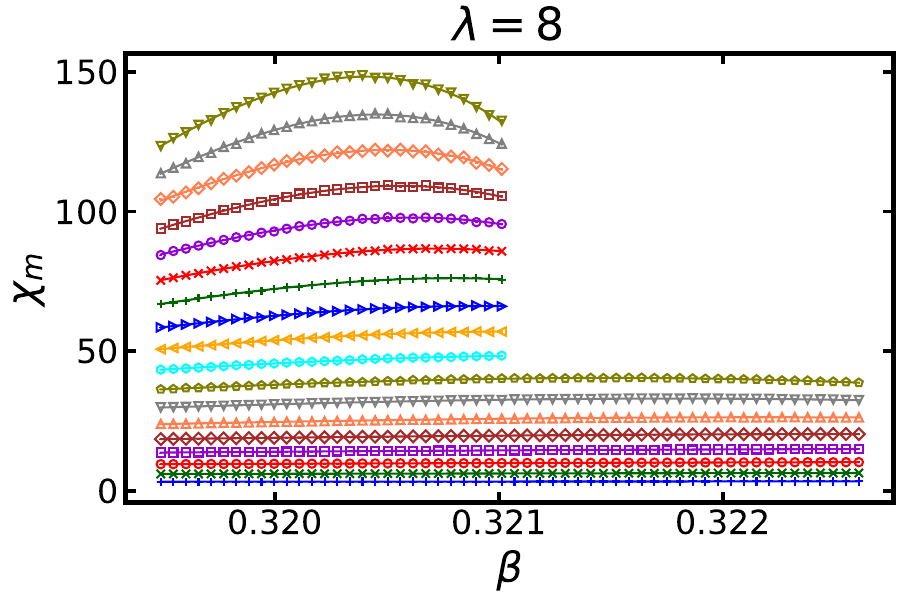}
    \caption{\label{fig_raw_chi}
    The raw data of the magnetic susceptibility $\chi_m$ are plotted against the temperature $\beta$.
    The left and right panels correspond to the results of the Heisenberg model $\lambda=0$ 
    and the local Heisenberg-dipolar model $\lambda=8$, respectively.
    The fitting results of the interpolating function~\eqref{eq_interpolate_chi} are also shown for each $L$.
    The correspondence between the colored symbols and $L$ in these plots is the same as that in Figure~\ref{fig_raw_U}.}
\end{figure*}

\begin{figure*}[tb]
    \centering
    \includegraphics[width=0.8\columnwidth]{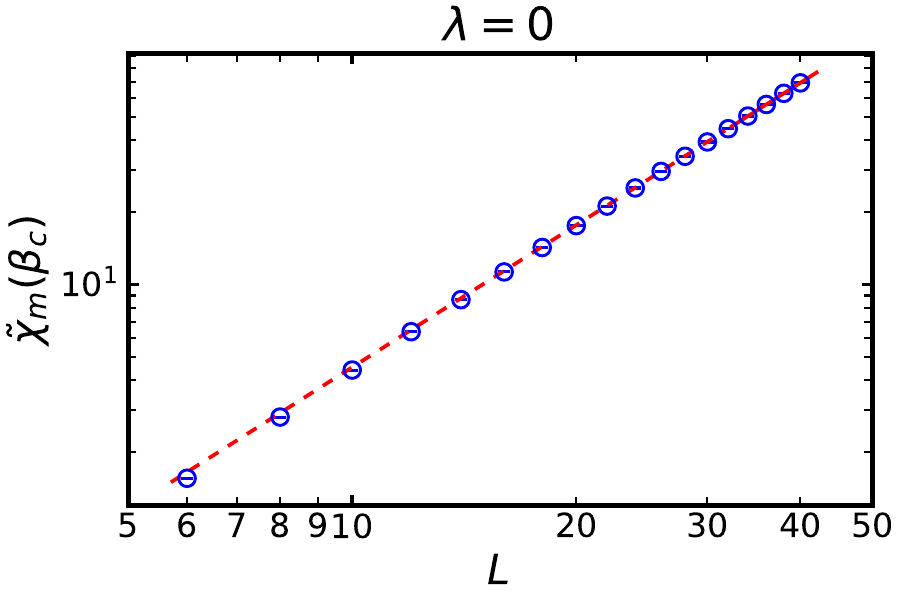}
    \quad
    \includegraphics[width=0.8\columnwidth]{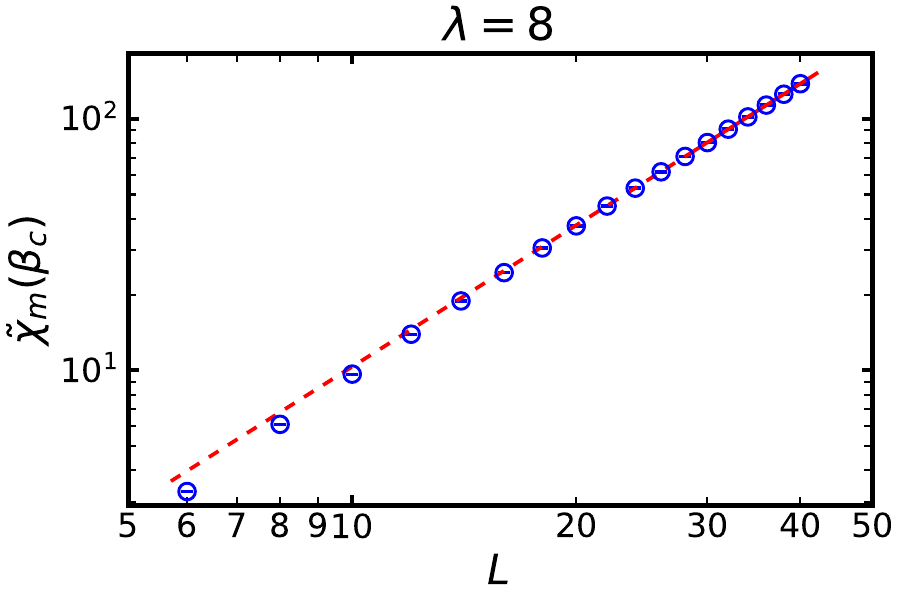}
    \caption{\label{fig_chic_vs_L}
    The $L$ dependence of the magnetic susceptibility at the critical temperature is plotted in log-log scale for $\lambda=0$ (left) and $8$ (right). The data is obtained by the interpolating function~\eqref{eq_interpolate_chi} as $\tilde{\chi}_m(\beta_c)|_L$.
    The fitting results by $b_0 L^{2-\eta}$ are also depicted by the solid (dashed) lines inside (outside) the fitting range, which is
    $L \in [32, 40]$ and $[28, 40]$ for $\lambda=0$ and $8$, respectively.}
\end{figure*}
Since we have already obtained the critical temperature $\beta_c$ as Eq.~\eqref{eq_beta_c}, 
we can compute $\chi_m$ at $\beta_c$ straightforwardly using the interpolating function~\eqref{eq_interpolate_chi}.
The resulting values of $\tilde{\chi}_m(\beta_c)|_L$ for each $L$ are shown in Figure~\ref{fig_chic_vs_L} in log-log scale, 
where the expected power-law behavior for $L$ is observed.

We fit the data points with a function $b_0 L^{2-\eta}$ of $L$ as shown in Figure~\ref{fig_chic_vs_L} 
to test the scaling behavior~\eqref{eq_scaling_chi}. 
Setting the fitting ranges to $L \in [32, 40]$ and $[28, 40]$, 
we obtained the best-fit values $(\eta, b_0) = (0.031(5), 0.0486(8))$ and $(0.131(8), 0.140(4))$ 
for $\lambda=0$ and $8$, respectively.
The fitting results are shown in Figure~\ref{fig_chic_vs_L}, 
where the data agree with the simple power-law scaling for sufficiently large $L$.

\begin{figure*}[tb]
    \centering
    \includegraphics[width=0.8\columnwidth]{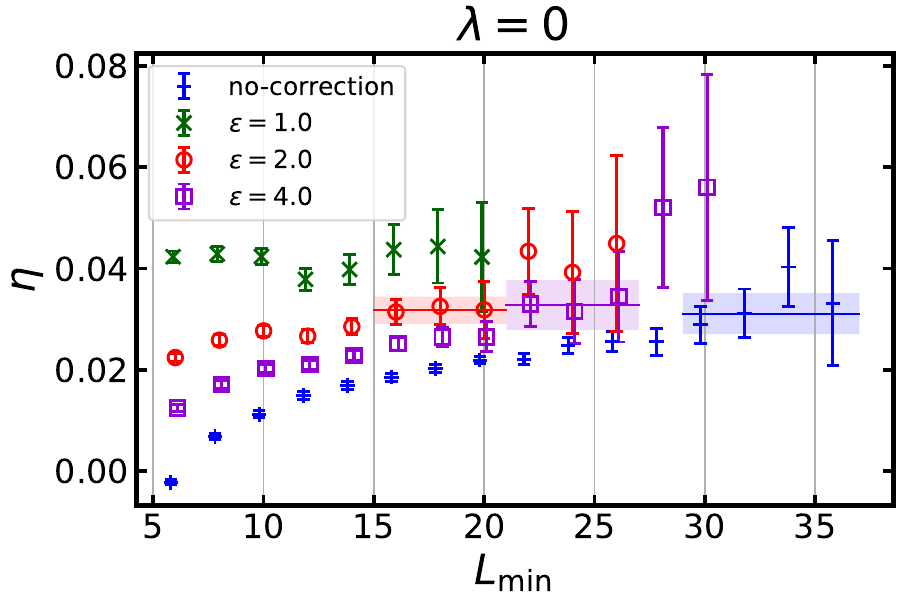}
    \quad
    \includegraphics[width=0.8\columnwidth]{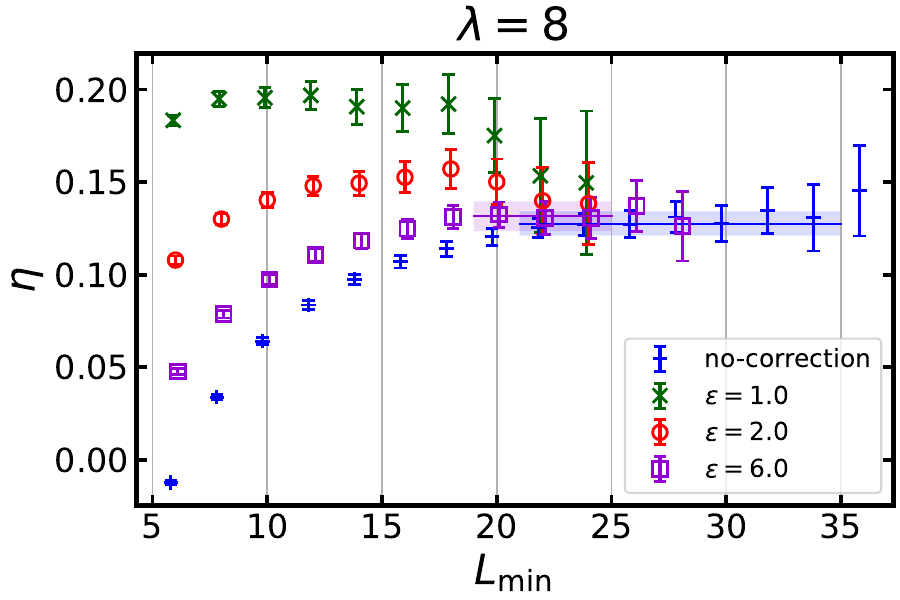}
    \caption{\label{fig_eta_vs_Lmin}
    The results of $\eta$ obtained via the fitting ansatz~\eqref{eq_fit_func_eta} are plotted 
    against the lower bound $L_{\mathrm{min}}$ of the fitting range for $\lambda=0$ (left) and $8$ (right).
    In both panels, the symbols of, plus-blue, cross-green, and circle-red depict 
    the results with $\epsilon=\infty$, $1.0$, and $2.0$, respectively.
    The square-purple symbols correspond to $\epsilon=4.0$ in the left panel and $6.0$ in the right panel.
    The colored horizontal lines depict the results of the constant fitting with the error bands.}
\end{figure*}
We estimate the systematic error from the finite size effect and determine the critical exponent $\eta$ more precisely.
Adding the finite size correction term to Eq.~\eqref{eq_scaling_chi}, we consider the scaling function to be
\begin{equation}
    \chi_{m,c}(L) = b_0 L^{2-\eta} (1 + b_1 L^{-\epsilon}),
    \label{eq_fit_func_eta}
\end{equation}
with the parameters $\eta$, $b_0$, and $b_1$.
As we did in the case of $\nu$ with Eq.~\eqref{eq_fit_func_nu}, 
we again set $\epsilon>0$ by hand to take the finite-size correction into account.
We fit the data points of $\tilde{\chi}_m(\beta_c)|_L$ by the no-correction ansatz ($\epsilon = \infty$) 
and the finite $\epsilon$ ansatz. 
Here, we present the results of $\eta$ against $L_{\mathrm{min}}$ with $\epsilon = 1.0, 2.0, 4.0$ for $\lambda = 0$ and $\epsilon = 1.0, 2.0, 6.0$ for $\lambda = 8$ in Figure~\ref{fig_eta_vs_Lmin}. In each panel, we can see that a plateau appears in large $L_{\mathrm{min}}$ regions.

For $\lambda=0$ with $\epsilon=2.0$ (circle-red symbols), $4.0$ (square-purple), and the no-correction ansatz (plus-blue),
we choose the plateau regions $L_{\mathrm{min}} \in [16, 20]$, $[22, 26]$, and $[30, 36]$ 
and then the constant fittings yield $\eta = 0.032(3)$, $0.033(5)$, and $0.031(4)$, respectively.
Note that the value of $\eta$ in the plateaus is relatively insensitive to the choice of $\epsilon$ values.
These results are depicted by the horizontal lines with the shadow bands in Figure~\ref{fig_eta_vs_Lmin} (left).

We take the result of $\epsilon=2.0$ as the central value and use the others to estimate the systematic errors from the uncertainty of $\varepsilon$, which results in 
\begin{equation}
    \eta = 0.0318(27) \left(^{+10}_{-7}\right) \left(^{+27}_{-28}\right)_{\beta_c} \qquad \text{for } \lambda = 0
    \label{eq_eta_result_lam0}
\end{equation}
for the Heisenberg model.
Here, the first, second, and third parentheses denote the statistical error, the systematic error from $\varepsilon$, and the systematic error from $\beta_c$, respectively.
For reference, the conformal bootstrap predicts $\eta = 0.0379(1)$~\cite{Chester:2020iyt}.
Here, the error in the bootstrap bound is not one-sigma range but rigorous. 

As for $\lambda=8$, we choose the range $L_{\mathrm{min}} \in [20, 24]$ with $\epsilon=6.0$ (square-purple) 
and obtain the central value, $\eta = 0.132(8)$, by the constant fitting in the range $L_{\mathrm{min}} \in [20, 24]$.
To estimate the systematic error, we fit the result of the no-correction ansatz (plus-blue) 
in the region $L_{\mathrm{min}} \in [22, 34]$, resulting in $\eta = 0.128(7)$.
These results are depicted by the horizontal lines with the shadow bands in Figure~\ref{fig_eta_vs_Lmin} (right).
Then our result of the exponent $\eta$ is given by 
\begin{equation}
    \eta = 0.132(8) \left(^{+0}_{-4}\right) \left(^{+12}_{-11}\right)_{\beta_c} \qquad \text{for } \lambda = 8
    \label{eq_eta_result_lam8}
\end{equation}
for the local Heisenberg-dipolar model. 

Finally, the critical exponent $\gamma$ is given by the relation $\gamma=\nu(2-\eta)$ 
combining the results of $\nu$ and $\eta$ obtained above:
\begin{equation}\label{eq:result_gamma}
    \gamma = 
    \begin{cases}
        1.4009(32) \left(^{+58}_{-0}\right)_{\nu} \left(^{+5}_{-7}\right)_{\eta} \left(^{+5}_{-7}\right)_{\beta_c} & (\lambda=0), \\
        1.123(6) \left(^{+0}_{-3}\right)_{\nu} \left(^{+2}_{-0}\right)_{\eta} \left(^{+16}_{-14}\right)_{\beta_c} & (\lambda=8).
    \end{cases}
\end{equation}
Here, the first parentheses represent the statistical error while the second, third, and fourth denote the systematic errors from $\nu$, $\eta$, and $\beta_c$, respectively.

\begin{figure}[htbp]
    \centering
    \includegraphics[width=0.8\columnwidth]{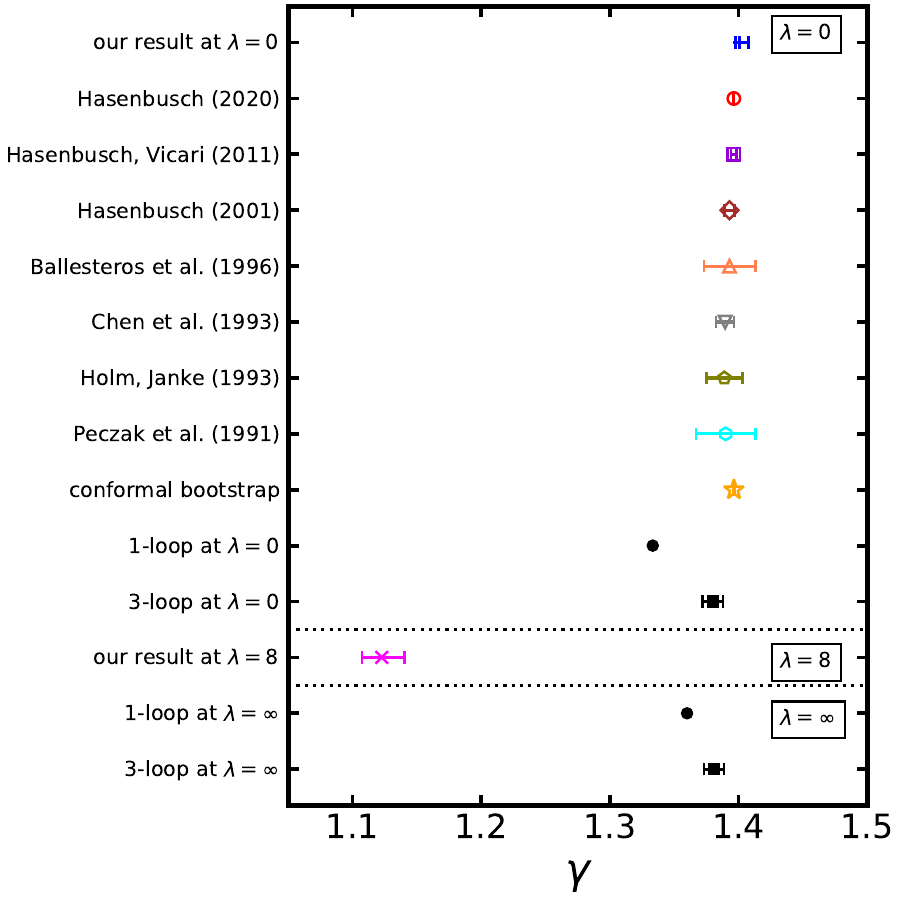}
    \caption{\label{fig_gamma_comp}
    The critical exponent $\gamma$ obtained in this paper and the previous studies 
    by the Monte Carlo~\cite{
    Hasenbusch:2020pwj, Hasenbusch:2011zwv, Hasenbusch:2000ph, 
    Ballesteros:1996bd, Chen:1993zz, Holm:1993zz, Peczak:1991zz} (colored open symbols), 
    the conformal bootstrap~\cite{Chester:2020iyt} (star-yellow), 
    the $\varepsilon$-expansion~\cite{2022NuPhB.98515990K} (filled symbols) are compared.
    The filled circle-black symbols depict the 1-loop results of the $\varepsilon$-expansion in Ref.~\cite{Aharony_1} 
    and the filled square-circle ones are the 3-loop results in Ref.~\cite{2022NuPhB.98515990K}.}
\end{figure}

Our results of $\gamma$ are compared with the results of the previous Monte Carlo studies~\cite{
Hasenbusch:2020pwj, Hasenbusch:2011zwv, Hasenbusch:2000ph, 
Ballesteros:1996bd, Chen:1993zz, Holm:1993zz, Peczak:1991zz}, 
the $\varepsilon$-expansion~\cite{Aharony_1, 2022NuPhB.98515990K}, 
and conformal bootstrap~\cite{Chester:2020iyt} in Figure~\ref{fig_gamma_comp}.
The 3-loop calculations yield coincidental results $\gamma=1.380(8)$ and $1.381(8)$ 
for the Heisenberg model ($\lambda=0$) and the Heisenberg-dipolar model with $\lambda=\infty$~\cite{Kudlis_2022, Gimenez-Grau:2023lpz}.
Also, the conformal bootstrap predicts $\gamma = 1.3964(5)$~\cite{Chester:2020iyt}.
On the other hand, our numerical results of $\gamma$ show a clear difference between $\lambda=0$ and $8$.
This observation suggests a possibility that the difference between the critical exponents at the dipolar and Heisenberg fixed points 
is more significant than the expectation from the 3-loop calculation.
Although our numerical method has room to be improved for more rigorous investigation, 
our result is consistent with the previous studies at least for $\lambda=0$.
For instance, the latest one, $\gamma=1.39635(20)$, obtained by the Monte Carlo study~\cite{Hasenbusch:2020pwj} 
is $1.4\sigma$ consistent(see the detailed comparison in Figure~\ref{fig_gamma_comp}).

\section{Correlation functions and transverse suppression }
\label{sec_cf_result}
In this section, we investigate the correlation function at the critical temperature.
Since the simulation is performed on a finite lattice, we set the temperature to $\beta = \beta_{\mathrm{peak}}$ 
where the magnetic susceptibility $\chi_m$ takes the maximum value, instead of $\beta_c$ of the thermodynamic limit.
We determine $\beta_{\mathrm{peak}}$ by quadratic-function fitting of the data $\chi_m$ only around the peak.
Now, we fix the lattice size to $L=40$ and obtain 
$\beta_{\mathrm{peak}} = 0.690190(12)$, $0.352883(11)$, and $0.320374(6)$ for $\lambda=0$, $4$, and $8$, respectively. 
The detail of the fitting is explained in Appendix~\ref{sec_chi_peak}.

After obtained the values of $\beta_{\mathrm{peak}}$, we generated $1,600$ configurations 
at intervals of $10^4$, $5 \times 10^4$, and $10^5$ sweeps 
for $\lambda=0$, $4$, and $8$, respectively, taking into account the autocorrelation.
We measure the connected part of the two-point correlation function in the $x$-direction, 
\begin{equation}
    \ev*{S_i(\vec{0})\, S_j(\vec{x})}_{\mathrm{conn}} 
    = \ev*{S_i(\vec{0})\, S_j(\vec{x})} - \ev*{S_i(\vec{0})} \ev{S_j(\vec{x})},
\end{equation}
where $i,j \in \{x,y,z\}$ label each component of the spin $\vec{S}$.
In this work, we take $\vec{x}=(x,0,0)$ and compute
\begin{equation}
    C_{ij}(\vec{x}) := \frac{1}{L^2}\sum_{y,z} \ev*{S_i(0,y,z)\, S_j(x,y,z)}_{\mathrm{conn}} .
\end{equation}

\begin{figure*}[tb]
    \centering
    \includegraphics[width=1.7\columnwidth]{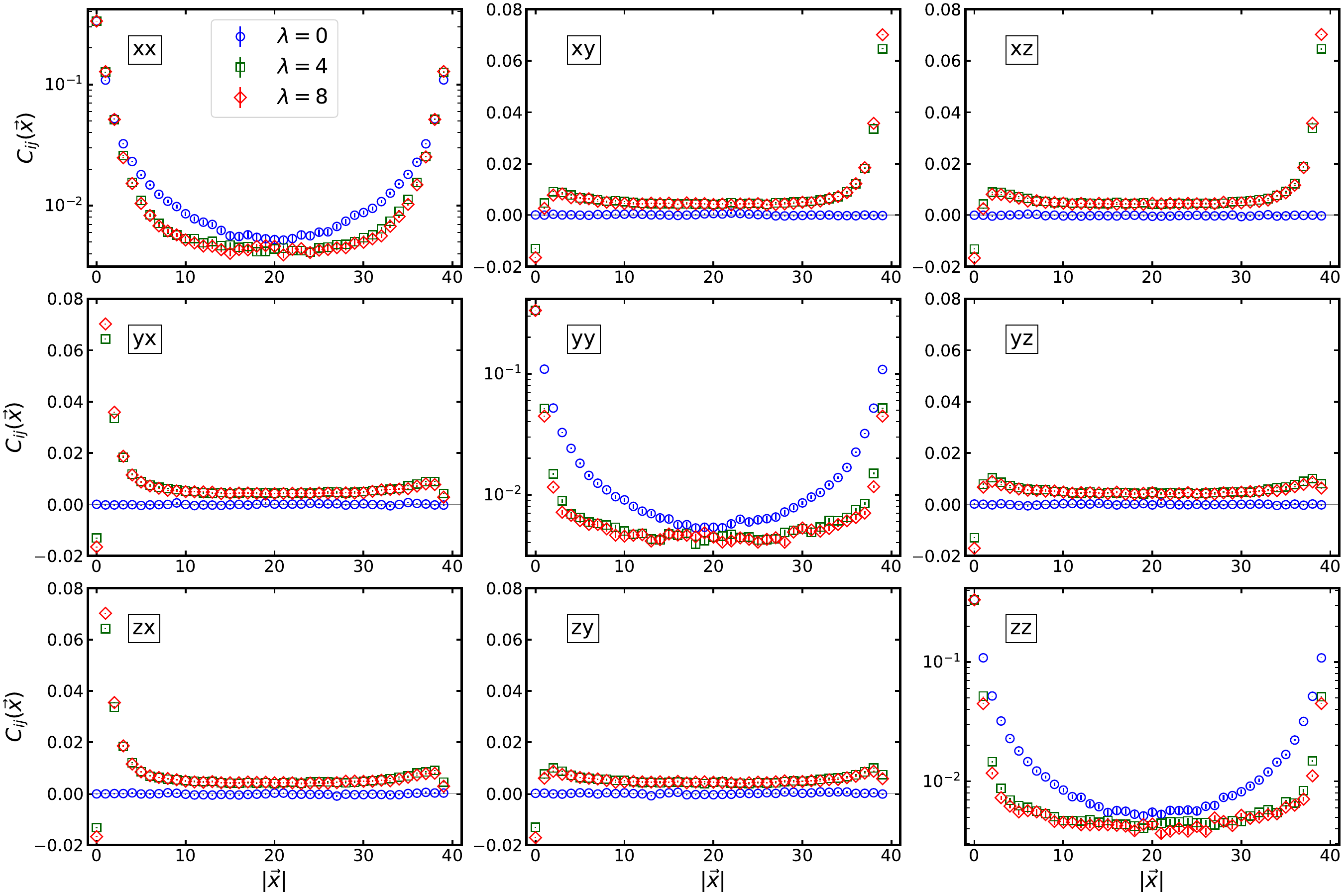}
    \caption{\label{fig_cf_matrix}
    The nine elements of the correlation function $C_{ij}(\vec{x})$ measure on the $x$-axis $\vec{x}=(x,0,0)$ 
    are plotted against the lattice coordinate $x$.
    The circle-blue, square-green, and diamond-red symbols correspond to $\lambda=0$, $4$, and $8$, respectively.
    The lattice size is fixed to $L=40$.
    Note that the diagonal elements are plotted on the semi-log scale whereas the off-diagonal ones are on the linear scale.}
\end{figure*}

The raw data of the correlation function are shown in Figure~\ref{fig_cf_matrix}, 
where the $3 \times 3$ elements are arranged like a matrix form.
For the Heisenberg model with $\lambda=0$ (circle-blue symbol), 
all the diagonal elements are equivalent within the statistical error, 
leaving the off-diagonal elements zero. 
This is expected from the $O(3)_S \times O(3)_L$ symmetry of the model: Here $O(3)_S$ acts on spin index and exact while $O(3)_L$ is the space rotation symmetry that is broken due to the lattice structure.
On the other hand, for the local Heisenberg-dipolar model with $\lambda=4$ (square-green) and $\lambda=8$ (diamond-red), 
we observe a clear difference between the $xx$ element and the others, 
which results from the symmetry-breaking into $O(3)_L$ due to the dipolar interaction, 
as expected in Eq.~\eqref{eq:corr-fn}. This $O(3)_L$ is further broken due to the lattice structure.

Let us compare this result with the field theory shown in Section~\ref{sec_theory}.
As explained, the two-point function \eqref{eq:corr-fn} is proportional to $\qty(\delta_{ij} - \alpha\frac{x_i x_j}{x^2})$. 
If we take $\vec{x}=(x,0,0)$, 
then the second term has a spin-component dependence, resulting in the total ratio of $(1-\alpha)$ for $i=j=x$ and $1$ for $i=j=y$ ( or $z$). For small $\eta$, $\alpha$ is negative, so the field theory predicts the suppression of the transverse direction (i.e. $yy$ and $zz$) compared with the longitudinal one (i.e. $xx$). This is indeed seen in our numerical plots.  

On the other hand, we also find the off-diagonal elements take nonzero values, in contrast with the two-point function~\eqref{eq:corr_conf} in the continuum field theory. This is due to the anisotropy caused by the finite-size effect.  For instance, the periodicity in lattice implies effective separation in $xy$ or $xz$ direction in addition to the intended $x$ separation. Moreover, the lack of parity invariance in the off-diagonal component comes from the forward derivative in our local dipolar-constraint term, Eq.~\eqref{eq_diff_lattice}.

\begin{figure*}[tb]
    \centering
    \includegraphics[width=1.7\columnwidth]{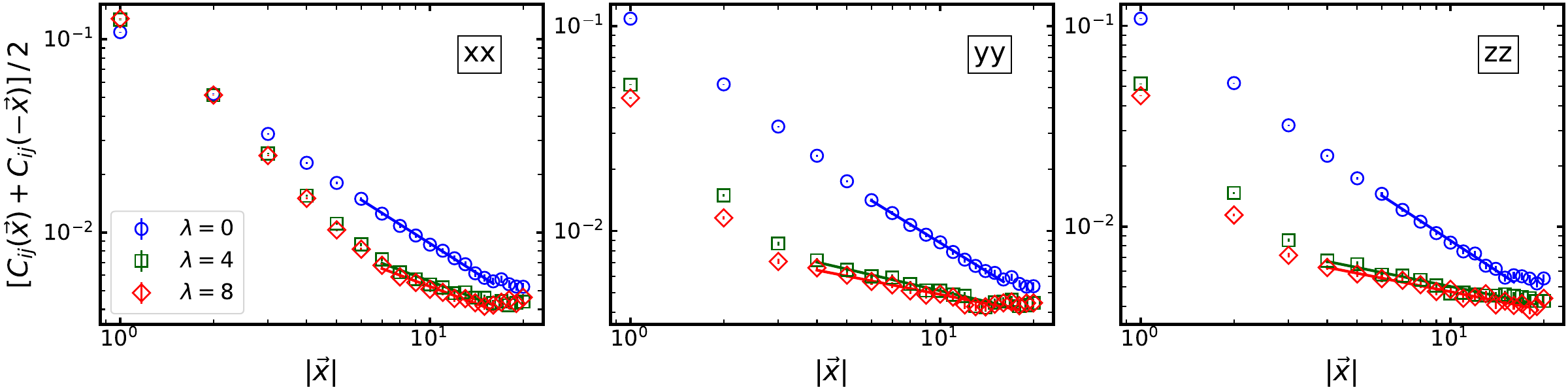}
    \caption{\label{fig_cf_fit}
    The forward-backward averages $[C_{ij}(\vec{x})+C_{ij}(-\vec{x})]/2$ of the diagonal elements 
    of the correlation function are plotted against $|\vec{x}|$ in log-log scale.
    The left, center, and right panels show the $xx$, $yy$, and $zz$ elements, 
    where the circle-blue, square-green, and diamond-red symbols correspond to $\lambda=0$, $4$, and $8$, respectively.
    The solid lines depict the fitting results by $c/|\vec{x}|^p$.}
\end{figure*}

Next, we estimate the critical exponent related to the anomalous dimension from the correlation functions.
We fit the diagonal elements of $C_{ij}(\vec{x})$, assuming a power function $c/|\vec{x}|^p$ in long $|\vec{x}|$ regimes.
Since the diagonal elements are symmetric under $\vec{x} \to -\vec{x}$, 
we take the forward-backward average, $[C_{ij}(\vec{x})+C_{ij}(-\vec{x})]/2$, before fitting.
The fitting range is chosen as a region where the power-law behavior is observed.
For $\lambda=0$, we set the range to $|\vec{x}| \in [6, 16]$ for all the diagonal elements.
As for $\lambda=4$ and $8$, the range is $|\vec{x}| \in [7, 16]$ for the $xx$ element and $|\vec{x}| \in [4, 18]$ for the others.
The fitting results are summarized in Table~\ref{tab_fit_cf_L040} 
and plotted in Figure~\ref{fig_cf_fit} with the data of the forward-backward average.
\begin{table}[h]
    \centering 
    \begin{tabular}{|c|c|c|c|c|c|}
        \hline $\lambda$ & element & $p$ & $c$ & $\chi^2/\mathrm{dof}$ & fit range \tabularnewline
        \hline \hline  
          & $xx$ & 1.02(4) & 0.0910(67) & 0.31 & $[6, 16]$ \tabularnewline
        \cline{2-6}
        0 & $yy$ & 0.93(4) & 0.0741(54) & 0.28 & $[6, 16]$ \tabularnewline
        \cline{2-6}
          & $zz$ & 1.01(4) & 0.0872(71) & 0.94 & $[6, 16]$ \tabularnewline
        \hline \hline  
          & $xx$ & 0.60(5) & 0.0223(22) & 1.61 & $[7, 16]$ \tabularnewline
        \cline{2-6}
        4 & $yy$ & 0.35(3) & 0.0114(6) & 1.01 & $[4, 18]$ \tabularnewline
        \cline{2-6}
          & $zz$ & 0.31(3) & 0.0103(6) & 0.95 & $[4, 18]$ \tabularnewline
        \hline \hline  
          & $xx$ & 0.60(5) & 0.0208(23)  & 0.98 & $[7, 16]$ \tabularnewline
        \cline{2-6}
        8 & $yy$ & 0.30(2) & 0.0097(5) & 0.87 & $[4, 18]$ \tabularnewline
        \cline{2-6}
          & $zz$ & 0.31(3) & 0.0096(5) & 0.51 & $[4, 18]$ \tabularnewline
        \hline
    \end{tabular}
    \caption{\label{tab_fit_cf_L040}
    The best-fit values of $p$ and $c$ obtained by fitting the correlation with $c/|\vec{x}|^p$ are summarized.
    The lattice size is set to $L=40$.}
\end{table}

In the case of the Heisenberg model ($\lambda = 0$), we find $p \approx 1$ identically for the three elements, as expected. 
Thus, we have $\Delta_\phi \approx 1/2$ in Eq.~\eqref{eq:def-alpha}. 
On the other hand, at $\lambda=8$ we find that the value of $p$ for $C_{yy}(\vec{x})$ and $C_{zz}(\vec{x})$ 
are consistent with each other, while $C_{xx}(\vec{x})$ is different from them.
Furthermore, both values are clearly different from the one for the Heisenberg model, 
thus it indicates that the model with $\lambda=0$ and $8$ are governed by a different scaling law.

The power $p = 2\Delta_{\phi}$ is related to the critical exponent $\eta$ via Eq.~\eqref{eq_delta_and_eta} 
as $\Delta_{\phi}=(1+\eta)/2$ with $d=3$.
Indeed, we obtain $\Delta_\phi \approx 1/2$ from the correlation function for $\lambda=0$, 
which is consistent with $\eta = 0.0318(27)(^{+10}_{-7})(^{+27}_{-28})_{\beta_c}$
in Eq.~\eqref{eq_eta_result_lam0} from the magnetic susceptibility as it is in the same order as the fitting error of $p$.
As for $\lambda=8$, we have $\Delta_{\phi} \approx 0.3$ from the $xx$ component,\footnote{
We could have read $\Delta_\phi$ from $yy$ and $zz$ component, which gives different values here. 
At two-loop, the anomalous dimensions of $xx$ and $yy$ (or $zz$) components are different for finite $\lambda$ 
(the difference becomes zero in the $\lambda \to \infty$ limit).
This may be a potential cause of the difference observed here.} which implies a negative value of $\eta$.
However, we obtained the positive value $\eta = 0.132(8)(^{+0}_{-4})(^{+12}_{-11})_{\beta_c}$ 
in Eq.~\eqref{eq_eta_result_lam8} from the magnetic susceptibility.
Thus, the two results obtained from the different observables appear inconsistent in the case of the local Heisenberg-dipolar model.
Note that, in reflection positive theory such as the Heisenberg model, it is strictly proved that $\eta$ must be positive.
On the other hand, this restriction does not apply to the local Heisenberg-dipolar model since it violates the reflection positivity.
In addition, even in the continuum theory, it is known that the derivative interaction can cause a negative anomalous dimension for the scalar field theory~\cite{Higashijima:2003et}.
We cannot exclude the possibility of the negative $\eta$ at the dipolar fixed point.

Let us make some remarks on this analysis.
First, to see the dipolar fixed point we have to consider the extrapolated value of $\eta$ in the $\lambda\to\infty$ limit. See Appendix ~\ref{sec_scaling_for_lambda} for a detailed analysis.
The next comments are subtleties in the analysis of the correlation function we have observed.
The power-law scaling region for $\lambda=8$ is relatively unclear than the case of $\lambda=0$.
It is possible that the finite-size effect on the correlation function becomes more significant for $\lambda=8$ and affects the power-law behavior.
Indeed, we did see some hints of this behavior by comparing the results for $L=32$ and $40$ in Appendix~\ref{sec_l_dep_cf}.
There, we see that the power $p$ read from the correlation functions of different lattice sizes differs: 
the larger the lattice size $L$, we had the larger $p$. 
It is possible that larger $L$ gives a consistent estimate of $\eta$ with the other method.
Furthermore, we observed that the correlation function is sensitive to the temperature, and thus it could be the source of ambiguity as well.
Here we choose the peak temperature $\beta_{\mathrm{peak}}$ of the magnetic susceptibility, 
but the behavior of the correlation function changes even slightly by shifting the temperature.
We examine the temperature dependence in Appendix~\ref{sec_off_peak_cf}.

All in all, we find reading critical exponents from the correlation function is less accurate than reading them from the finite size corrections of Binder parameters and magnetic susceptibility, both in the Heisenberg model and the local Heisenberg-dipolar model. This is in accord with what has been observed in the literature. We therefore report the values in Section~\ref{sec_result} as our main results.

\section{Summary and discussion}
\label{sec_summary}
In this paper, we investigated the critical exponents and correlation functions for the local Heisenberg-dipolar model using the Monte Carlo simulation.
It is motivated by the discussion given by Aharony and Fisher that non-local dipolar effects in magnetism destabilize the Heisenberg fixed point, leading to a new fixed point, called the dipolar fixed point. To effectively simulate the fixed point, we introduced the local lattice Hamiltonian by adding a local dipolar constraint term to that of the Heisenberg model, with the coupling denoted as $\lambda$. In the limit $\lambda \rightarrow \infty$, we expect that it coincides with the dipolar fixed point of Aharony and Fisher, which belongs to a different universality class than the Heisenberg fixed point.

Our simulation results reproduced the critical exponents of the Heisenberg model reported in previous studies when $\lambda = 0$. On the other hand, for $\lambda = 8$, the behavior was clearly different from that of the Heisenberg model, strongly suggesting that the dipolar fixed point has distinct critical exponents.

Furthermore, examining the spin component dependence of the two-point correlation function, we found that strong transverse suppression occurs for $\lambda = 4$ and $8$, reflecting the dipolar constraint. 
The original $O(3)_S \times  O(3)_L$ symmetry in the Heisenberg model is broken down into $O(3)_L$ symmetry for $\lambda \ne 0$. Our results reproduced the properties. In the limit of $\lambda \rightarrow \infty$, that is, at the dipolar fixed point, it is expected that the theory remains scale-invariant but is no longer conformally invariant.

Our obtained value of the critical exponent $\gamma$ turned out to be smaller than the estimation of the other method. There might exist a theoretical, as well as experimental, interpretation of this apparent smallness (if we assumed that it is indeed different from the dipolar fixed point value $\gamma_{\mathrm{dipolar}}$)~\footnote{The following interpretation was communicated to us by Slava Rychkov. We would like to thank him for his insight.} .
Recall that our simulation at finite $\lambda$ does not correspond to a true RG fixed point but rather to a crossover point, which has been pointed out by K.~Ried et al.~\cite{Ried:1995} (see also Ref.~\cite{aharony202450}). 
In such cases, it is argued that the effective susceptibility exponent $\gamma_{\mathrm{eff}} = \frac{d \log \chi }{d \log \beta}$ may show a significant dip away from the fixed point~\cite{Gimenez-Grau:2023lpz}. 
In our case, the deviation from the RG fixed point is mainly caused by finite $\lambda$, and our definition of $\gamma$ is different from $\gamma_{\mathrm{eff}}$, but it is plausible that the apparent smallness of our obtained $\gamma$ may be attributed to the similar effects. We also note that there is experimental evidence for such an effect in dipolar ferromagnets (see e.g. \cite{inbook} and reference therein).

Therefore, one important future direction is to clarify the $\lambda$ dependence of critical exponents and estimate these values at the dipolar fixed point of Aharony and Fisher.
An extrapolation focusing on $B=\langle (\nabla \cdot \vec{S})^2 \rangle$ instead of $\lambda \rightarrow \infty$ as shown in Appendix ~\ref{sec_scaling_for_lambda} might be a realistic approach in these future analyses. 

Another important future problem is to consider the thermodynamic limit.
As discussed, our results up to $L=40$ may still have a large finite-size effect. 
To do that, the improvement of simulation algorithms would be necessary.
In this paper, we used the simplest Metropolis algorithm and encountered severe autocorrelation around the critical temperature, 
namely the critical slowing down, which prevented us from taking a larger lattice size than $L=40$, in particular with larger $\lambda$.
For more precise determination of the critical exponents, 
the cluster algorithm~\cite{PhysRevLett.58.86, PhysRevLett.62.361} can deal with this problem 
since it efficiently reduces the autocorrelation by a non-local update of the spin configuration.
However, it is not straightforward to apply the algorithm to the local Heisenberg-dipolar model 
due to the next-nearest-neighbor interaction, the first term in the second line of Eq.~\eqref{eq_H_by_M}.
In this case, the generalized versions of the cluster algorithm will be helpful~\cite{PhysRevLett.65.941, PhysRevB.43.8539}.
Alternatively, it is also interesting to simulate the long-range dipolar interaction directly without using the local description.
Indeed, the efficient algorithms for spin models with non-local interaction are proposed 
in Refs.~\cite{doi:10.1142/S0129183195000265, Fukui_2009}.

We have replaced the transverse (or dipolar) constraint on the local magnetization vector with the local cost function in the Hamiltonian. There are some pros and cons to this replacement. The obvious pro is it is easier to implement as the Monte Carlo simulation with the local Hamiltonian. The major con is that the constraint term has a renormalization group fixed point only at $\lambda = \infty$ and the finite $\lambda$ does not give a fixed point in a strict sense. Even worse, the fixed point at $\lambda = \infty$ is unstable, meaning that the correction cannot be neglected in the large size limit (formally $\omega$ is negative). This may explain our observation that the scaling behavior with finite $\lambda$ is less precise than the one at $\lambda = 0$. The necessity of taking the infinite $\lambda$ limit is challenging in the current formulation because the auto-correlation time becomes larger with larger $\lambda$. We are, however, optimistic: some new ideas presented in the previous paragraph will pave the way.

Finally, we wish there would be a continuous effort to experimentally verify the dipolar effects in real ferromagnets~\footnote{In the Chinese character, the root of magnetism is associated with a stone with benevolence. Why? In ancient Chinese stories, it is said that the way the ferromagnet attracts iron rocks makes it look as if an infant is naturally attracted to his or her mother. Probably the dipolar magnet will embody more benevolence because the magnetism from it is divergenceless, so it appears to be more gentle and embracing. So far in human history, the distinct features of dipolar magnetism have been observed in esoteric materials such as Europium compounds, but the universal nature of the benevolence should also be seen in more conventional materials such as iron or nickel in future experiments, just with more care.}.

\acknowledgments
The numerical calculations were carried out on Yukawa-21 at YITP in Kyoto University and the PC clusters at RIKEN iTHEMS.
The work of E.~I. is supported by JST PRESTO Grant Number JPMJPR2113, 
JSPS Grant-in-Aid for Transformative Research Areas (A) JP21H05190, 
JSPS Grant Number JP23H05439, 
JST Grant Number JPMJPF2221,  
JPMJCR24I3,  
and also supported by Program for Promoting Researches on the Supercomputer ``Fugaku'' (Simulation for basic science: from fundamental laws of particles to creation of nuclei) and (Simulation for basic science: approaching the new quantum era), and Joint Institute for Computational Fundamental Science (JICFuS), Grant Number JPMXP1020230411. 
The work by YN is in part supported by JSPS KAKENHI Grant Number 21K03581. 
This work is supported by Center for Gravitational Physics and Quantum Information (CGPQI) at Yukawa Institute for Theoretical Physics.

\appendix

\section{Renormalization group study of the local Heisenberg-dipolar model}
\label{sec:beta-fn}
Within the $\varepsilon$ expansion at one-loop, the $\beta$ functions are given by 
\begin{equation}
    \begin{aligned}
        \beta_t &= 2t+u\qty[C\qty(1-\frac{\lambda}{2(1+2\lambda)})\Lambda_0^2 + tC\qty(-1 +\frac{\lambda(1+\lambda)}{(1+2\lambda)^2})], \\
        \beta_u &= \varepsilon u - \frac{u^2C}{3}\qty[ \frac{6+18\lambda + 17\lambda^2}{(1+ 2\lambda)^2}], \\
        \beta_\lambda &= 0, 
    \end{aligned}
\end{equation}
where $\Lambda_0$ is the ultraviolet cutoff and $C$ is $\frac{S^4}{(2\pi)^4}=\frac{1}{8\pi^2}$. 
Corresondingy, the non-trivial fixed points are locted at
\begin{equation}
    \begin{aligned}
        t^{\star} &= -\frac{3\varepsilon (1+2\lambda)^2}{2(6 + 13\lambda + 17\lambda^2)}\qty(1 - \frac{\lambda}{2(1+2\lambda)})\Lambda^2, \\
        u^{\star} &= \frac{3\varepsilon(1+2\lambda)^2}{C(6+13\lambda + 17\lambda^2)}.
    \end{aligned}
\end{equation}
By linearizing the beta functions around the fixed point, we obtain the critical exponent $\nu$ shown in Eq.~\eqref{eq:1-loop}.

\section{Hubbard-Stratonovich transformation}
\label{sect_HS_trsf}
Mapping of the parameters in the continuum field theory of Eq.~\eqref{eq:local_discription} and a lattice model in Section~\ref{sec_lattice_formula} is non-trivial.
Specifically, in Eq.~\eqref{eq:local_discription}, the term that imposes an important constraint in the local Heisenberg-dipolar model may acquire a multiplicative factor when mapped to the lattice model at finite $\lambda$.  We derive the factor from the Hubbard-Stratonovich transformation within the mean field approximation.

Let us start the Hubbard–Stratonovich transformation in the more general setup.
We consider the partition function given by the bilinear of spin variables:
\begin{equation}
    Z = \Tr\exp\qty(\frac{\beta}{2}\sum_{n,m \in \Lambda} \vec{S}_n \cdot K_{nm} \vec{S}_m 
    + \beta\sum_{n\in\Lambda} \vec{h}_n \cdot \vec{\phi}_n),
    \label{eq_org_partitional}
\end{equation}
where $\Lambda$ represents a set of lattice sites of the size $|\Lambda|=\sum_{n\in\Lambda}1$ 
and the matrix $K$ is assumed to be positive.
Introducing the auxiliary variable vector $\phi_n$, we insert the identity, 
\begin{align}
    1 &= \qty(\frac{\beta}{2\pi})^{|\Lambda|/2}\frac{1}{\sqrt{\det K}}\notag \\
    &\quad \times \int \qty(\prod_{l\in\Lambda}d\phi_l)
    \exp\qty(-\frac{\beta}{2}\sum_{n,m\in\Lambda}\vec{\phi}_n \cdot K^{-1}_{nm}  \vec{\phi}_m),
\end{align}
into Eq.~\eqref{eq_org_partitional} and perform the change of variable $\phi_n \to \phi_n - \sum_m K_{nm} S_m$.
We obtain 
\begin{align}
    Z & = C\prod_{n\in\Lambda} \int d\phi_n
    \Tr\exp\left\{-\frac{\beta}{2}\left[ \sum_{n,m \in \Lambda} \vec{\phi}_n \cdot K_{nm}^{-1} \vec{\phi}_m\right.\right. \notag \\
    &\quad\quad \left.\left. -\sum_{n\in \Lambda} 2(\vec{\phi}_n + \vec{h}_n ) \cdot \vec{S}_n \right]\right\},
    \label{eq_befor_tr} 
\end{align}
where $C = \qty(\frac{\beta}{2\pi})^{|\Lambda|/2}\frac{1}{\sqrt{\det K}}$.

Since the spin variable $S$ only appears linearly, one can evaluate the trace over $S$ independently at each site. Up to quadratic orders in $\phi$, the resulting expression is
\begin{align}
Z
&\sim \prod_{n=1}^{|\Lambda|} \int_{-\infty}^\infty d\phi_n 
\exp\left\{-\frac{\beta}{2}\sum_{n,m}\vec{\phi}_n \cdot (K^{-1})_{nm}\vec{\phi}_m\right. \notag\\
&\quad\quad\left. + \frac{\pi\beta^2 \mu^2}{3}\sum_n (\vec{\phi}_n + \vec{h}_n)^2\right\}.
\end{align}
Finally, by taking the continuum limit, we can rewrite the lattice partition function~Eq.~\eqref{eq_org_partitional} into the continuum one:
\begin{align}
    Z &\sim \int \mathcal{D}\vec{\phi} \exp\left(-\frac{\beta}{2}\int_{x,y}\vec{\phi}(x) \cdot K^{-1}(x - y)\vec{\phi}(y) \right. \notag \\
    &\quad\quad \left.+\frac{\pi\beta^2 \mu^2}{3} \int_x(\vec{\phi}(x)+\vec{h})^2\right),
    \label{eq_cont_model}   
\end{align}
where $\vec{\phi}$ denotes a three-component vector field, $\phi_i(x)$.

Now, we specify our lattice Hamiltonian based on the Heisenberg model, adding the local dipolar constraint term on the lattice:
\begin{equation}
    H = -\frac{J}{2}\sum_{n,i}\qty(\vec{S}_n - \vec{S}_{n+\hat{i}})^2 
    + \lambda\sum_n(\nabla \cdot\vec{S}_n )^2 + \frac{1}{2}J\theta\sum_n \vec{S}\,^2,
    \label{eq_lattice_model_original}
\end{equation}
where $\hat{i}\in\{\hat{1},\hat{2},\hat{3}\}$ denotes the lattice unit vector in the $i$-th direction.
We have added the last term (with $\theta >6$) as a trick because we want to make the matrix $K$ positive.

In the continuum limit, we identify the quadratic form as
\begin{align}
    &\int d^3xd^3y\ \vec{\phi}(x) \cdot \tilde{K}(x-y) \vec{\phi}(y) \notag\\
    &= \int d^3x \qty(-J\theta \phi_i \phi_i  + Ja^2(\partial_i\phi_j)^2 - 2a^2\lambda(\partial_i\phi_i)^2 ), 
    \label{eq_K}
\end{align}
where $a$ is lattice size spacing (we will set $a=1$).
Taking the inverse of $\tilde{K}$, 
\begin{align}
        &\int d^3xd^3y\ \vec{\phi}(x) K^{-1}(x-y) \vec{\phi}(y) \notag\\
        &\simeq \int d^3x \qty(-\frac{1}{J\theta}\phi_i\phi_i - \frac{a^2}{J\theta^2} (\partial_i\phi_j)^2 + \frac{2a^2\lambda}{J^2\theta^2}(\partial_i\phi_i)^2).
\end{align}
Here, we assume $\phi$ as varying slowly in space in the IR region, and then $(I - \epsilon A)^{-1} \simeq I + \epsilon A$. 
Thus, the Hamiltonian finally becomes 
\onecolumngrid
\begin{align}
\label{eq_3.10}
    H &= \int d^3x\ \qty(-\frac{1}{2J\theta}\phi_i\phi_i - \frac{a^2}{2J\theta^2}(\partial_i\phi_j)^2 + \frac{a^2\lambda}{J^2\theta^2}(\partial_i\phi_i)^2 
  -\frac{\pi\beta\mu^2}{6}({\phi_i}(x)+{h}_i)^2).
\end{align}
\twocolumngrid
Consequently, we show that the normalization of $\lambda$ here is the one introduced in section~\ref{sec_theory} when we take $J=-1$ in the lattice Hamiltonian~\eqref{eq_lattice_model_calc} as a corresponding model.

\section{Determination of the peak position of \texorpdfstring{$\chi_m$}{chim}}
\label{sec_chi_peak}
\begin{figure*}[tb]
    \centering
    \includegraphics[width=0.6\columnwidth]{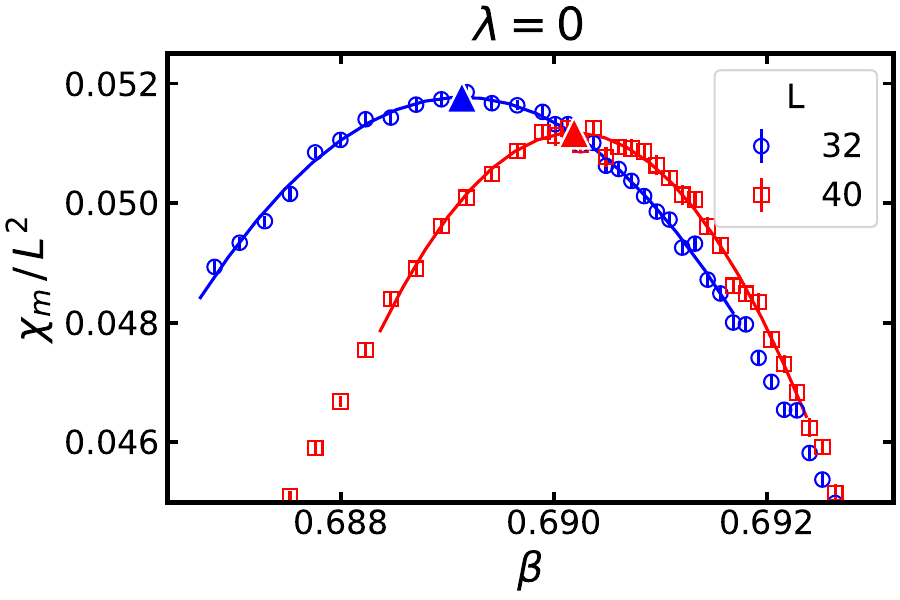}
    \includegraphics[width=0.6\columnwidth]{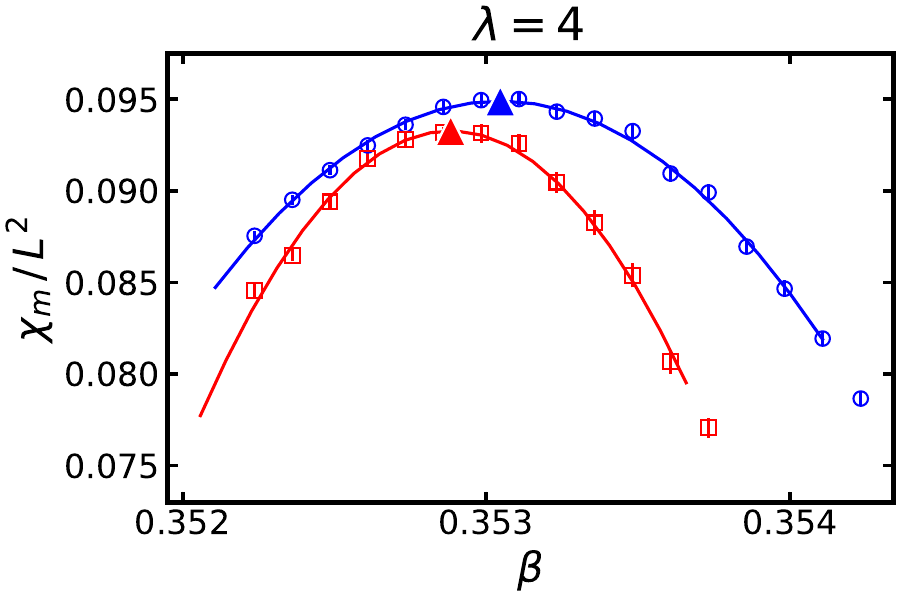}
    \includegraphics[width=0.6\columnwidth]{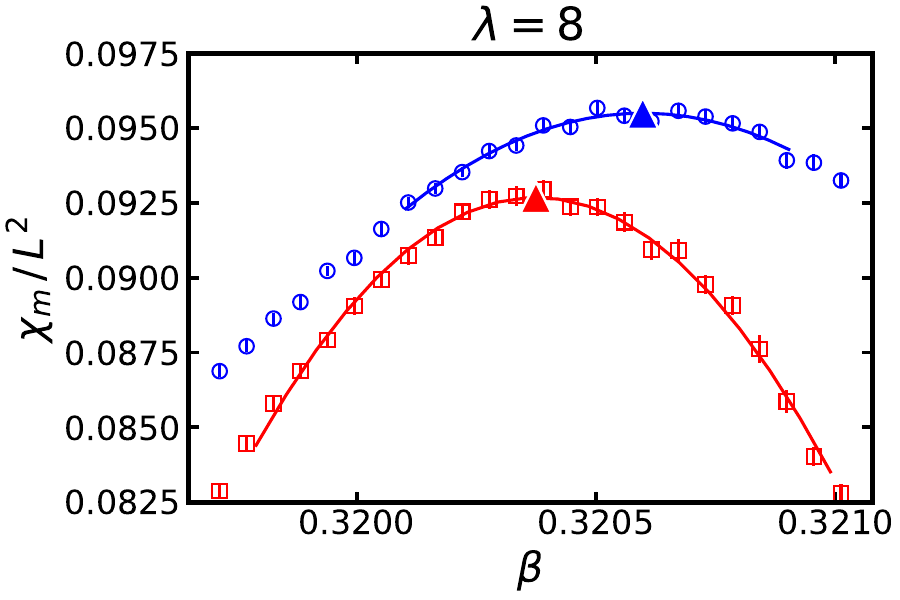}
    \caption{\label{fig_chi_peak}
    The fitting results of the magnetic susceptibility $\chi_m$ are shown, 
    where the peak positions are determined to measure the correlation function.
    The left, center, and right panels correspond to $\lambda=0$, $4$, and $8$, respectively.
    The peak positions are depicted by the triangle symbols on the fitting curves.
    To make the figures easier to see, we plot the value of $\chi_m$ divided by $L^2$.}
\end{figure*}

We measure the correlation function at the peak position of the magnetic susceptibility $\chi_m$.
To determine the peak position, we fit the data points of $\chi_m$ around the peak by the quadratic function.
Here the lattice size is set to $L=32$ or $40$.
The fitting curves are displayed in Figure~\ref{fig_chi_peak}, where the triangle symbols denote the peak positions $\beta_{\mathrm{peak}}$.

In Table~\ref{tab_beta_peak}, the explicit values of $\beta_{\mathrm{peak}}$ are summarized with the results of $\beta_c$ in Eq.~\eqref{eq_beta_c}.
We can see that $\beta_{\mathrm{peak}}$ approaches $\beta_c$ as $L$ increases.
\begin{table}[h]
    \centering 
    \begin{tabular}{|c|c|c|c|}
        \hline
         & $\lambda=0$ & $\lambda=4$ & $\lambda=8$ \tabularnewline
        \hline \hline
        $\beta_{\mathrm{peak}}^{(L=32)}$ & $0.689135(17)$ & $0.353046(8)$ & $0.320598(12)$ \tabularnewline
        \hline
        $\beta_{\mathrm{peak}}^{(L=40)}$ & $0.690190(12)$ & $0.352883(11)$ & $0.320374(6)$ \tabularnewline
        \hline \hline
        $\beta_c$ & $0.693035(11)(^{+18}_{-20})$ & $-$ & $0.319844(18)(^{+43}_{-41})$ \tabularnewline
        \hline
    \end{tabular}
    \caption{\label{tab_beta_peak}
    The peak positions $\beta_{\mathrm{peak}}$ of the magnetic susceptibility $\chi_m$ obtained by the quadratic fitting are summarized.
    We also put the results of $\beta_c$ in Eq.~\eqref{eq_beta_c} for comparison.}
\end{table}

\section{Fitting results of the interpolating functions}
\label{sec_fit_result_interpolate}
Here, we summarize the best-fit values of the parameters $\{c_i(L)\}$ and $\{d_i(L)\}$ 
of the interpolating functions~\eqref{eq_interpolate_U} and~\eqref{eq_interpolate_chi} 
which are obtained in Sections~\ref{subsec_intersecting_binder} and~\ref{subsec_exponent_gamma} in Table~\ref{tab_c_and_d}.
The fitting errors of these values are large in some cases 
because the $\chi^2$-fit suffers from a relatively flat direction in the parameter spaces of $\{c_i(L)\}$ and $\{d_i(L)\}$.
Even though the fitting parameters have large errors, the interpolated values have smaller errors, used to obtain the intersection and gradient, etc.
Thus, it does not cause a problem in our analyses.

\begin{table*}
    \centering
    $\lambda = 0$ 
    {\footnotesize \include{fig_lam0/table_c_and_d.tex}}
    $\lambda = 8$ 
    {\footnotesize \include{fig_lam8/table_c_and_d.tex}}
    \caption{\label{tab_c_and_d}
    The best-fit values of the parameters $\{c_i(L)\}$ and $\{d_i(L)\}$ 
    in the interpolating function~\eqref{eq_interpolate_U} and~\eqref{eq_interpolate_chi} 
    for $\lambda=0$ (top) and $8$ (bottom).
    These values are normalized by $L^2$ or $L^4 {10}^3$ to be almost the same order in these tables.}
\end{table*}

\section{
Scaling Behavior in the \texorpdfstring{$\lambda \to \infty$ limit }{lambda to infinity}
}
\label{sec_scaling_for_lambda}
For the Heisenberg-dipolar model, the dipolar constraint is completely imposed in the $\lambda\to\infty$ limit.
In this appendix, we test the validity of this extrapolation.
As we measured the correlation function at $\lambda=0$, $4$, and $8$, 
it is feasible to perform the three-point extrapolation to $\lambda\to\infty$, 
assuming no phase transition at the finite $\lambda$.
However, a naive extrapolation by a power-function of $1/\lambda$ cannot apply to $\lambda=0$.
Therefore, we use $B = \sum_n B_n / V$ instead of $\lambda$ as a parameter of the dipolar constraint, 
where $B_n = \ev*{(\vec{\nabla} \cdot \vec{S}_n)^2}$ is defined in Section~\ref{subsec_validity_dipolar}.
Since $B$ is finite at $\lambda = 0$ and vanishes in the $\lambda\to\infty$ limit, the extrapolation to $B \to 0$ is well-defined.

\begin{figure}[h]
    \centering
    \includegraphics[width=0.8\columnwidth]{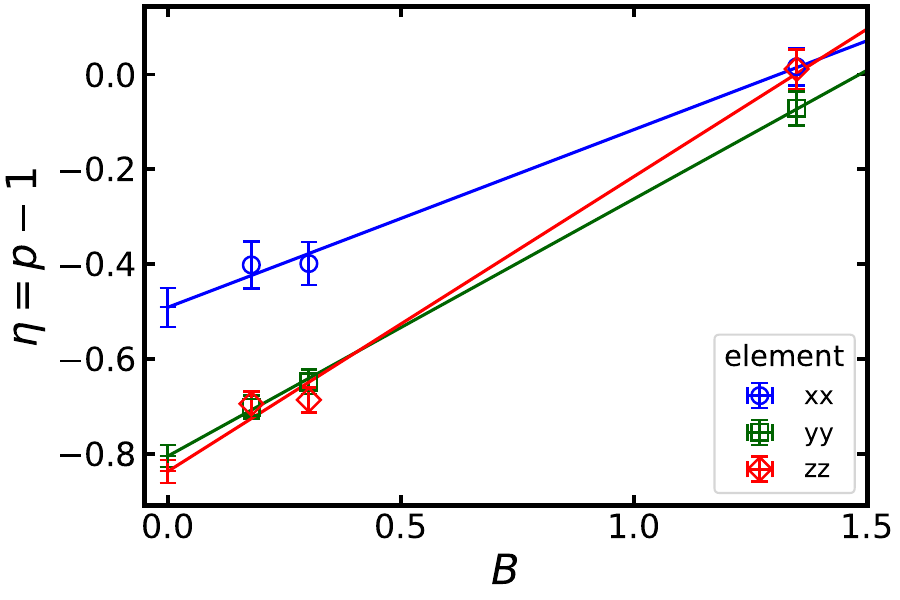}
    \caption{\label{fig_eta_vs_B}
    The critical exponent $\eta$ obtained from each diagonal element of the correlation function 
    is plotted against $B = \sum_n B_n / V$ for the lattice size $L=40$.
    The symbols of circle-blue, square-green, and diamond-red correspond to the result of the $xx$, $yy$, and $zz$ elements, respectively.
    The horizontal error bar of $B$ is invisibly small.
    The fitting lines of $\eta = \eta_0 + aB$ and the extrapolated values are depicted by the solid lines and plus symbols, respectively.}
\end{figure}

In Figure~\ref{fig_eta_vs_B}, the critical exponent $\eta$ obtained from the correlation function 
is plotted against $B$ for the three cases of $\lambda$, 
where $\eta = p-1$ is taken from Table~\ref{tab_fit_cf_L040} for each element of the correlation function at $L=40$.
The data points show almost linear behavior, implying $\eta$ and $B$ obey the same scaling for $\lambda$.
Therefore, we fit the data points by a linear function, $\eta = \eta_0 + aB$, 
and obtain $(\eta_0, a) = (-0.49(4), 0.37(5))$, $(-0.81(2), 0.54(4))$, and $(-0.84(2), 0.62(4))$ 
for the $xx$, $yy$, and $zz$ elements, respectively.
Note that the extrapolated value $\eta_0$ is the result at the finite lattice size $L$ and is sensitive to $L$.
Furthermore, as pointed out in Refs.~\cite{Ried:1995, aharony202450, inbook}, the $\lambda$ dependence of the exponents might be nonmonotonic.
Increasing the number of data points for $\lambda$ would reveal the entire behavior of this model.

\section{Finite size effect on the correlation function}
\label{sec_l_dep_cf}
\begin{figure*}[tb]
    \centering
    \includegraphics[width=1.7\columnwidth]{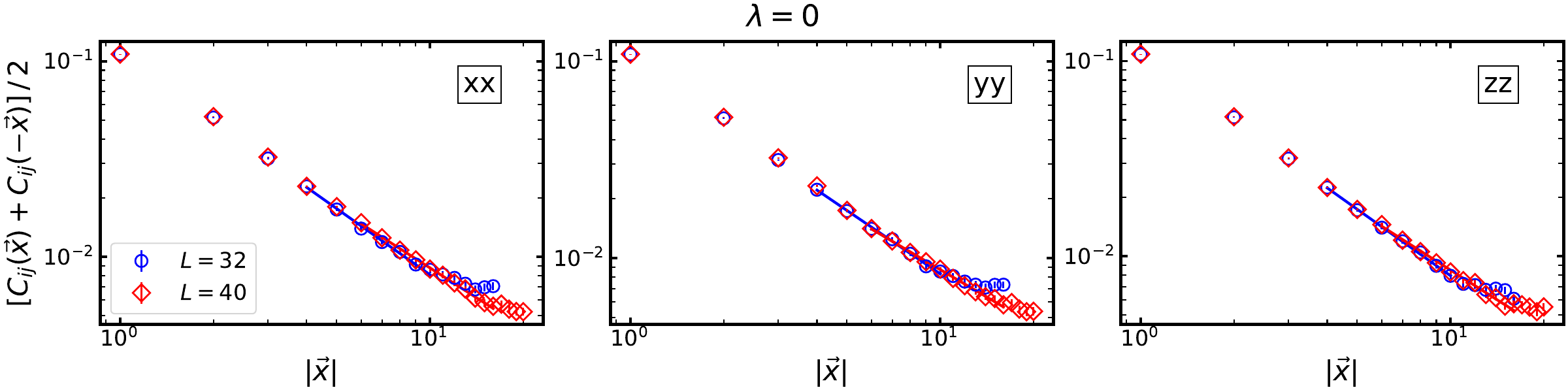}
    \includegraphics[width=1.7\columnwidth]{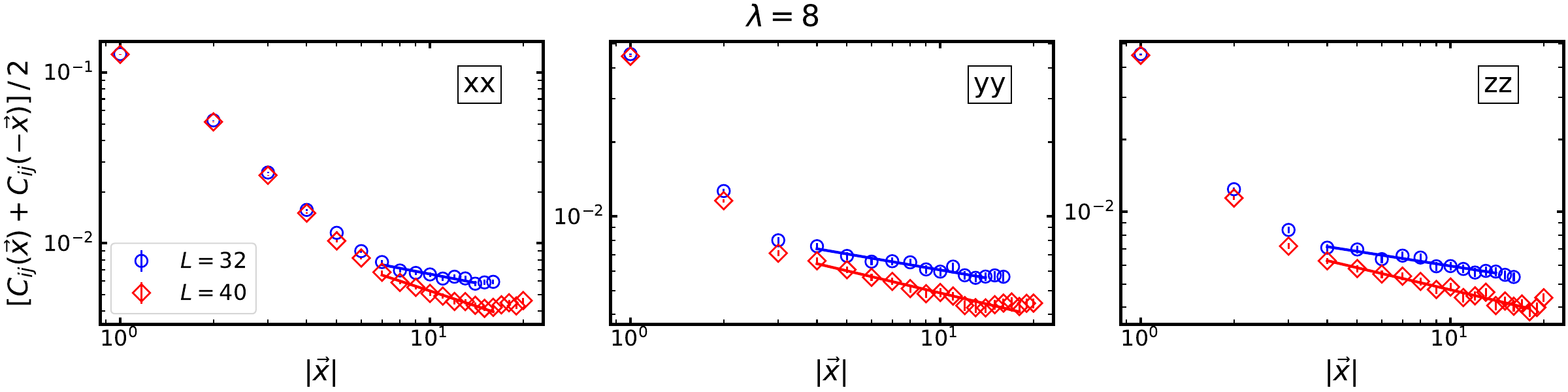}
    \caption{\label{fig_cf_l_dep}
    The forward-backward averages $[C_{ij}(\vec{x})+C_{ij}(-\vec{x})]/2$ of the diagonal elements 
    of the correlation function are plotted against $|\vec{x}|$ in log-log scale for $\lambda=0$ (top) and $8$ (bottom).
    The left, center, and right columns show the $xx$, $yy$, and $zz$ elements, 
    where the circle-blue and diamond-red symbols correspond to $L=32$ and $40$, respectively.
    The solid lines depict the fitting results by $c/|\vec{x}|^p$.}
\end{figure*}

Here, we discuss the system-size dependence of the correlation function.
In Figure~\ref{fig_cf_l_dep}, the correlation functions at the different system sizes $L=32$ and $40$ are compared, 
where the forward-backward average $[C_{ij}(\vec{x})+C_{ij}(-\vec{x})]/2$ is taken.
The square-red data denote the result at the peak temperatures $\beta_{\mathrm{peak}}$ of the magnetic susceptibility on $L=40$ lattice, previously shown in Section~\ref{sec_cf_result}, 
The circle-blue data are on $L=32$ lattice, where we tune the value of $\beta$ as $\beta_{\mathrm{peak}}=0.689135$ and $0.320598$ for $\lambda=0$ and $8$, respectively.
We also depict the fitting lines of $c/|\vec{x}|^p$ in Figure~\ref{fig_cf_l_dep}, 
where the best-fit values are shown in Table~\ref{tab_fit_cf_L032} and~\ref{tab_fit_cf_L040} for $L=32$ and $40$, respectively.
\begin{table}[h]
    \centering 
    \begin{tabular}{|c|c|c|c|c|c|}
        \hline $\lambda$ & element & $p$ & $c$ & $\chi^2/\mathrm{dof}$ & fit range \tabularnewline
        \hline \hline  
          & $xx$ & 1.12(3) & 0.107(5) & 1.32 & $[4, 10]$ \tabularnewline
        \cline{2-6}
        0 & $yy$ & 1.07(3) & 0.098(4) & 0.66 & $[4, 10]$ \tabularnewline
        \cline{2-6}
          & $zz$ & 1.13(4) & 0.107(6) & 0.25 & $[4, 10]$ \tabularnewline
        \hline \hline  
          & $xx$ & 0.35(6) & 0.0149(19) & 0.99 & $[7, 16]$ \tabularnewline
        \cline{2-6}
        8 & $yy$ & 0.22(3) & 0.0100(6) & 0.49 & $[4, 18]$ \tabularnewline
        \cline{2-6}
          & $zz$ & 0.20(4) & 0.0095(7) & 0.38 & $[4, 18]$ \tabularnewline
        \hline
    \end{tabular}
    \caption{\label{tab_fit_cf_L032}
    The best-fit values of $p$ and $c$ obtained by fitting the correlation with $c/|\vec{x}|^p$ are summarized.
    The lattice size is set to $L=32$.}
\end{table}

In the case of the Heisenberg model ($\lambda=0$), the power-law scaling regions are observed for both $L=32$ and $40$.
Then we find $p \sim 1$ for both cases with a difference of about $10\%$ which could come from the finite-size effect.
On the other hand, in the case of the local Heisenberg-dipolar model ($\lambda=8$), 
we observe larger differences in the long-distance behavior between $L=32$ and $40$.
Indeed, the resulting values of $p$ depend on $L$ more than the case of $\lambda=0$, 
which indicates the presence of a significant finite-size effect in the case of $\lambda=8$.

In the case of $\lambda=8$, the best-fit values of $p$ at $L=40$ in Table~\ref{tab_fit_cf_L040} 
are larger than those at $L=32$ in Table~\ref{tab_fit_cf_L032}.
Thus, if we take the thermodynamic limit $L\to\infty$, 
there is a chance to get extrapolated values satisfying $p>1$, 
which corresponds to a positive value of the critical exponent $\eta$.
Although we mentioned the inconsistency between the values of $\eta$ 
estimated from the magnetic susceptibility and correlation function in Section~\ref{sec_cf_result}, 
it might be resolved by taking the thermodynamic limit in future research.

\section{Correlation function at off-peak temperatures}
\label{sec_off_peak_cf}
We examine the correlation functions at slightly off-peak temperatures 
around $\beta_{\mathrm{peak}}$ in Table~\ref{tab_beta_peak} to see how the determination of $\beta_{\mathrm{peal}}$ affects the correlation function and critical exponent.
Here we set $\lambda=4$, $L=32$ and choose two temperatures $\beta=0.352140$ and $0.353952$, 
where the value of the magnetic susceptibility $\chi_m$ becomes $0.9$ times the maximum one at $\beta_{\mathrm{peak}}=0.353046$.

\begin{figure*}[tb]
    \centering
    \includegraphics[width=1.7\columnwidth]{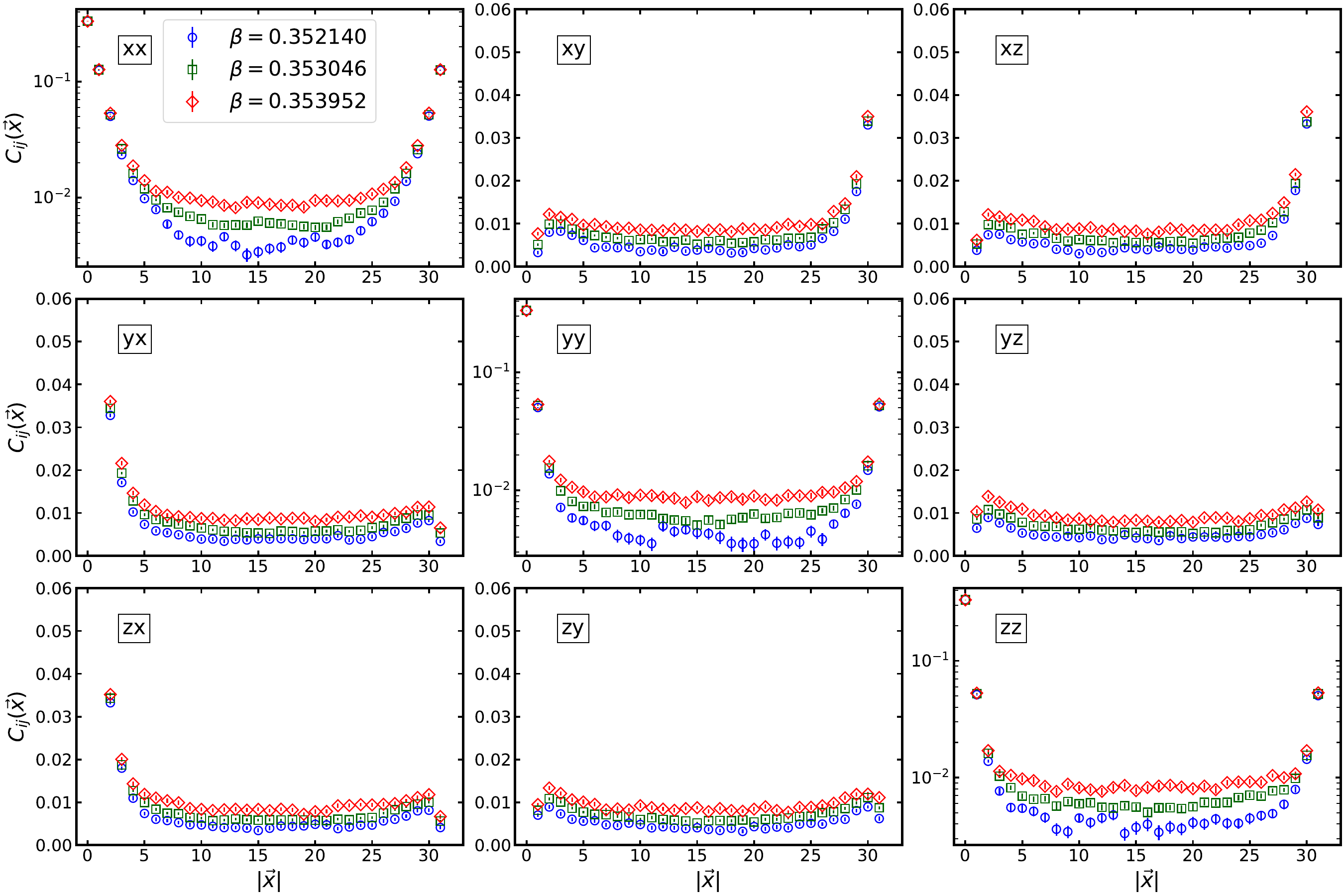}
    \caption{\label{fig_cf_off_peak}
    The nine elements of the correlation function are plotted against the lattice coordinate $x$ for $\lambda=4$ and $L=32$.
    The results obtained at the slightly off-peak positions ($\beta=0.352140$ and $0.353952$) 
    are depicted by the circle-blue and diamond-red symbols 
    whereas the result just at the peak position ($\beta_{\mathrm{peak}}=0.353046$) is represented by the square-green symbol.
    The diagonal elements are plotted on the semi-log scale and the off-diagonal ones are on the linear scale.}
\end{figure*}

We compare the correlation functions at the two off-peak temperatures with one at the peak position in Figure~\ref{fig_cf_off_peak}.
We can see that the behavior of the correlation function, particularly for the diagonal components, is significantly dependent on the temperature. 
Here, as $\beta$ increases toward the ordered phase, the correlation function becomes flat at long distances.
In our present simulation, it is hard to determine the value of $\beta_{\mathrm{peak}}$ more precisely, but it suggests that the value of $p$ in the present analysis has a sizeable systematic uncertainty.

\clearpage
\bibliography{biblio.bib}

\end{document}

%% file: fig_lam0/table_c_and_d.tex
\begin{tabular}{|c|c|c|c|c|c|c|c|}
\hline
$L$ & $c_0/L^2$ & $c_1/L^2$ & $c_2/L^2$ & $d_0/(L^4{10}^3)$ & $d_1/(L^4{10}^3)$ & $d_2/(L^4{10}^3)$ & $d_3/(L^4{10}^3)$ \tabularnewline
\hline
\hline
6 & 0.082(78) & -0.21(22) & 0.17(16) & 0.025(36) & -0.11(15) & 0.16(22) & -0.075(107) \tabularnewline
\hline
8 & -0.11(4) & 0.34(13) & -0.23(9) & 0.027(19) & -0.12(8) & 0.17(12) & -0.080(57) \tabularnewline
\hline
10 & -0.13(3) & 0.38(9) & -0.26(6) & -0.013(12) & 0.058(52) & -0.084(76) & 0.040(36) \tabularnewline
\hline
12 & -0.11(2) & 0.30(7) & -0.21(5) & -0.0021(94) & 0.0089(408) & -0.013(59) & 0.0060(283) \tabularnewline
\hline
14 & -0.19(2) & 0.53(5) & -0.37(3) & -0.00055(709) & 0.0024(307) & -0.0033(443) & 0.0016(213) \tabularnewline
\hline
16 & -0.21(1) & 0.59(4) & -0.42(3) & 0.0025(46) & -0.011(20) & 0.016(29) & -0.0076(138) \tabularnewline
\hline
18 & -0.23(1) & 0.65(3) & -0.46(2) & -0.0030(38) & 0.013(16) & -0.019(23) & 0.0089(113) \tabularnewline
\hline
20 & -0.23(1) & 0.67(2) & -0.47(2) & -0.0020(31) & 0.0085(136) & -0.012(20) & 0.0058(94) \tabularnewline
\hline
22 & -0.28(1) & 0.79(3) & -0.56(2) & -0.010(4) & 0.045(16) & -0.065(23) & 0.031(11) \tabularnewline
\hline
24 & -0.28(1) & 0.81(2) & -0.58(2) & -0.011(3) & 0.047(13) & -0.068(19) & 0.033(9) \tabularnewline
\hline
26 & -0.28(1) & 0.80(2) & -0.57(1) & -0.018(3) & 0.076(13) & -0.11(2) & 0.052(9) \tabularnewline
\hline
28 & -0.31(1) & 0.88(2) & -0.63(1) & -0.014(2) & 0.060(9) & -0.087(13) & 0.042(6) \tabularnewline
\hline
30 & -0.31(1) & 0.90(2) & -0.64(1) & -0.015(2) & 0.063(8) & -0.091(11) & 0.044(5) \tabularnewline
\hline
32 & -0.33(0) & 0.95(1) & -0.68(1) & -0.021(1) & 0.091(6) & -0.13(1) & 0.063(4) \tabularnewline
\hline
34 & -0.34(0) & 0.97(1) & -0.69(1) & -0.021(1) & 0.090(5) & -0.13(1) & 0.062(4) \tabularnewline
\hline
36 & -0.35(0) & 1.0(0) & -0.72(1) & -0.021(1) & 0.090(5) & -0.13(1) & 0.062(3) \tabularnewline
\hline
38 & -0.37(0) & 1.1(0) & -0.75(1) & -0.026(1) & 0.11(0) & -0.16(1) & 0.077(3) \tabularnewline
\hline
40 & -0.37(0) & 1.1(0) & -0.77(1) & -0.029(1) & 0.13(0) & -0.18(1) & 0.087(3) \tabularnewline
\hline
\end{tabular}

%% file: fig_lam8/table_c_and_d.tex
\begin{tabular}{|c|c|c|c|c|c|c|c|}
\hline
$L$ & $c_0/L^2$ & $c_1/L^2$ & $c_2/L^2$ & $d_0/(L^4{10}^3)$ & $d_1/(L^4{10}^3)$ & $d_2/(L^4{10}^3)$ & $d_3/(L^4{10}^3)$ \tabularnewline
\hline
\hline
6 & -0.065(128) & 0.36(79) & -0.37(123) & 0.010(20) & -0.097(190) & 0.30(59) & -0.32(61) \tabularnewline
\hline
8 & -0.0028(876) & -0.070(546) & 0.32(85) & 0.025(13) & -0.23(12) & 0.73(37) & -0.76(38) \tabularnewline
\hline
10 & -0.13(5) & 0.69(33) & -0.86(52) & -0.0023(77) & 0.021(72) & -0.063(225) & 0.063(234) \tabularnewline
\hline
12 & -0.11(3) & 0.56(21) & -0.67(33) & 0.0088(51) & -0.084(48) & 0.26(15) & -0.28(15) \tabularnewline
\hline
14 & -0.23(2) & 1.3(1) & -1.9(2) & 0.013(4) & -0.12(4) & 0.38(12) & -0.40(12) \tabularnewline
\hline
16 & -0.34(2) & 2.0(1) & -3.0(2) & 0.013(4) & -0.12(3) & 0.39(11) & -0.41(11) \tabularnewline
\hline
18 & -0.48(2) & 2.9(1) & -4.3(2) & 0.023(3) & -0.22(3) & 0.68(9) & -0.72(9) \tabularnewline
\hline
20 & -0.60(3) & 3.6(2) & -5.5(3) & 0.024(4) & -0.23(4) & 0.73(13) & -0.76(13) \tabularnewline
\hline
22 & -0.45(5) & 2.7(3) & -4.1(5) & 0.027(16) & -0.25(15) & 0.79(48) & -0.83(50) \tabularnewline
\hline
24 & -0.37(3) & 2.2(2) & -3.3(3) & 0.052(14) & -0.48(13) & 1.5(4) & -1.6(4) \tabularnewline
\hline
26 & -0.45(4) & 2.7(3) & -4.1(4) & 0.055(12) & -0.52(11) & 1.6(4) & -1.7(4) \tabularnewline
\hline
28 & -0.65(4) & 4.0(3) & -6.0(4) & 0.064(11) & -0.60(10) & 1.9(3) & -2.0(3) \tabularnewline
\hline
30 & -0.75(4) & 4.6(3) & -7.0(4) & 0.087(13) & -0.81(12) & 2.6(4) & -2.7(4) \tabularnewline
\hline
32 & -0.85(4) & 5.2(3) & -8.0(4) & 0.10(1) & -0.95(9) & 3.0(3) & -3.1(3) \tabularnewline
\hline
34 & -1.1(0) & 6.6(2) & -10(0) & 0.088(11) & -0.83(10) & 2.6(3) & -2.7(3) \tabularnewline
\hline
36 & -1.2(0) & 7.3(2) & -11(0) & 0.13(1) & -1.2(1) & 3.7(3) & -3.9(3) \tabularnewline
\hline
38 & -1.3(0) & 8.1(2) & -12(0) & 0.12(1) & -1.1(1) & 3.6(3) & -3.7(3) \tabularnewline
\hline
40 & -1.5(0) & 9.1(2) & -14(0) & 0.12(1) & -1.2(1) & 3.7(3) & -3.8(3) \tabularnewline
\hline
\end{tabular}